\documentclass[twocolumn, twocolappendix]{aastex701}

\shorttitle{WLM AGB SFH}
\shortauthors{Lee et al.}

\defcitealias{Lee25}{L25}
\defcitealias{Cohen25}{C25}

\begin{document}

\title{The Star Formation History of WLM from Asymptotic Giant Branch Stars and  \\The Discovery of a \added{Candidate} Accreted System in its Outer Disk}

\author[0000-0002-5865-0220]{Abigail~J.~Lee}\altaffiliation{NASA Hubble Fellow}\affil{Department of Astronomy, University of California, Berkeley, CA 94720-3411, USA}\email{abby.lee@berkeley.edu}

\author[0000-0002-6442-6030]{Daniel~R.~Weisz}\affil{Department of Astronomy, University of California, Berkeley, CA 94720-3411, USA}\email{dan.weisz@berkeley.edu}

\author[0000-0001-8416-4093]{Andrew~E.~Dolphin}\affil{Raytheon, 1151 E. Hermans Road, Tucson, AZ 85756, USA}\affil{Steward Observatory, University of Arizona, 933 N. Cherry Avenue, Tucson, AZ 85719, USA}\email{adolphin@rtx.com}

\author[0000-0002-1445-4877]{Alessandro~Savino}\affil{Department of Astronomy, University of California, Berkeley, CA 94720-3411, USA}\email{asavino@berkeley.edu}

\begin{abstract}

We measure the star formation history (SFH) of Local Group dwarf galaxy WLM using wide-area ($\sim4$~half-light~radii) ground-based near-infrared (NIR) imaging of bright ($M_{J}<-4.9$~mag) asymptotic giant branch (AGB) stars. From our NIR color-magnitude diagram (CMD) of 825~stars, we find that our recovered SFH is in excellent agreement with literature SFHs of WLM measured from much deeper CMDs ($M_{F090W}\sim+4.3$~mag) based on JWST imaging. We find good agreement in the qualitative shape of the SFHs as well as quantitative metrics such as the timescales for which 50\% and 90\% of the stellar mass formed with $\tau_{50,{\rm AGB}}=5.16_{-0.50}^{+2.07}$~Gyr ago and $\tau_{90,{\rm AGB}}=1.33_{-0.09}^{+0.11}$~Gyr ago versus $\tau_{50,{\rm JWST}}=5.94_{-0.31}^{+0.37}$~Gyr ago and $\tau_{90,~{\rm JWST}}=1.55_{-0.13}^{+0.03}$~Gyr ago. The coarser precision of the AGB star-based values is driven by the low number of AGB stars. We derive an age-metallicity relation (AMR) from the AGB star CMD fitting that is similar to the JWST-based AMR and is consistent with other reported metallicities in WLM (i.e., spectroscopy, RR Lyrae). From our wide-area AGB star map and SFH, we identify a stellar over-density ($M_*\sim2.0\times10^6~M_{\odot}$, $r_h\sim340$~pc) in WLM's northwestern outer disk. Despite WLM having long been considered an isolated galaxy, the mass, size, and age of this over-density are highly suggestive of an accreted dwarf galaxy. Overall, our findings illustrate the power of NIR observations of AGB stars for efficiently and accurately measuring SFHs and for identifying and characterizing substructures in nearby galaxies.

\end{abstract}

\correspondingauthor{Abigail J. Lee}\email{abby.lee@berkeley.edu}


\section{Introduction} 
\label{sec:intro}

Color-magnitude diagrams (CMDs) that extend below the oldest main sequence turnoff (oMSTO) are the `gold standard’ for measuring a galaxy's star formation history (SFH) across cosmic time \citep[e.g.,][]{Mateo98, Gallart05, Tolstoy09, Kennicutt12, Conroy13}.  Unfortunately, due to the crowding and faintness of the oMSTO ($M_V\sim+4$), even the Hubble Space Telescope (HST) has only been able to reach the oMSTO in galaxies within the Local Group (LG). While these SFHs measured from the oMSTO have had high-scientific impact, particularly on our knowledge of low-mass galaxies (e.g., \citealt{Harris04, Harris09, Cole07, Monelli10a, Monelli10b,Hidalgo11, Hidalgo13, Brown14, Skillman14,Skillman17, Weisz14a, Gallart15, Rubele15, Rubele18, Mazzi21, Sacchi21, Savino23, Savino25, Durbin25}), they have been inherently limited to the LG, which has a specific accretion history that may not be representative of the broader Universe. The James Webb Space Telescope (JWST) has produced the first CMD that reached the oMSTO just outside the LG (Leo P; $D\sim1.6$~Mpc; \citealt{McQuinn24b}).  But even with its excellent angular resolution and sensitivity, crowding and faintness will limit JWST observations of the oMSTO to distances of $\sim2-3$~Mpc.

Extending CMD-based SFHs to environments well-beyond the LG has relied and will continue to rely on evolved stars (e.g., \citealt{Gogarten10,Williams09, Williams10, McQuinn10a,McQuinn10b,  Weisz11,Cignoni18, Cignoni19, Sacchi18, Sacchi19, Smercina25}).  Typically, in the optical, CMD features such as the subgiant branch (SGB; $M_V\sim+3.5$), the horizontal branch (HB; $M_V\sim+0.5$), and red clump (RC; $M_V\sim+0.5$) are considered the next-best features on the CMD for measuring lifetime SFHs \citep[e.g.,][]{Gallart05}.  The oldest SGB is almost as faint as the oMSTO and provides only marginally better observational prospects \citep[e.g.,][]{Cole14}.  The HB is known to be highly sensitive to age \citep[e.g.,][]{Savino19}, but HB-based SFHs can be limited in precision by the small number of HB stars available and are highly dependent on the choice in stellar model \citep[e.g.,][]{Gallart05}. In principle, accurately modeling the HB on CMDs should include variations in more complex stellar physics (e.g., mass loss; \citealt{Savino19}). 

RC-depth CMDs have long been the target of resolved star observations in nearby galaxies as they provide a balance of realistic observational considerations and age-sensitivity (e.g., \citealt{Weisz08, Williams09, Williams17, McQuinn10a, Weisz11, Weisz14a, Smercina25}). In principle, the RC consists of core helium-burning stars with ages between $\sim1-10$ Gyr, which can provide age information. However, the effects of stellar evolution tend to skew the RC age distribution toward younger ($\sim1-4$ Gyr) and metal-rich populations \citep[e.g.,][]{Gallart05, Girardi16}.  It is also a phase of evolution known for its complexity; uncertainties in core overshooting, rotation, binarity, and mass loss  can be significant and are only marginally calibrated \citep[e.g.,][]{Girardi16}.  In practice, age information based on RC-depth CMDs is more tightly constrained by the ratio of RC to red giant branch (RGB) stars. While SFHs measured from RC-depth CMDs have been shown to be accurate, they only provide coarse time resolution at lookback times older than a few Gyr \citep[e.g.,][]{Weisz14a}.

An emerging alternative for measuring \added{CMD-based} SFHs in galaxies beyond the LG is asymptotic giant branch (AGB) stars. AGB stars are evolved intermediate and low-mass stars ($0.5-8M_{\odot}$) near the end of their lifetimes. 
\added{A review on the masses, ages, and different types of AGB stars can be found in Section 2 of \cite{Lee25}.}
AGB stars are known to be sensitive to age and are significantly brighter than other age-sensitive CMD features (e.g., \citealt{Frogel83,Reid84, Wood85, Costa96, Davidge03, Davidge05, Davidge14, Cioni05, Melbourne10, Jung12, Crnojevic13, Mcquinn17}). Their potential for measuring ages has been recognized for decades (e.g., \citealt{Gallart96, Cioni06a, Cioni06b, Cioni08, Held10, Javadi11b,Rejkuba22, Harmsen23, Velguth24}). 

However, two challenges have limited the practical use of AGB stars for \added{CMD-based} SFH measurements.  First, AGB stars are less age-sensitive in optical wavelengths \citep{  Javadi11a, Lee25}. Furthermore, $\sim40$\% of AGB stars, particularly the metal-rich and cool AGB stars, are undetected in optical wavelengths \citep{Zijlstra96, Jackson07,Boyer09, Boyer19, Held10}.  This ultimately has limited the utility of AGB star SFHs for the wealth of optical CMDs in the literature \citep[e.g.,][]{Holtzman06, Dalcanton09}.  Second, AGB stars exhibit a variety of complex physics (e.g., mass loss, dredge up, thermal pulses, self-extinction) that can limit the accuracy of age information \citep{Girardi10, Marigo17}, indicating that careful calibration of any age from information from AGB stars is required. 

Over the past decade, there have been two significant advances that changed the landscape for AGB star SFHs.
First, AGB stars ages in new stellar models developed in the past several years by the \texttt{COLIBRI} group \citep{Marigo17, Pastorelli19, Pastorelli20} have been anchored to SFHs measured from the oMSTO in the Magellanic Clouds \citep{Rubele18, Mazzi21}.  This places AGB star based ages on a self-consistent scale with MSTO stars when used in CMD modeling. Second, is the increasing abundance of near-infrared (NIR) imaging in nearby galaxies.  It is only at NIR wavelengths that both 
\added{(1) C-type and M-type AGB cleanly separate, and (2) TP-AGB stars are brighter than the TRGB}
(e.g., \citealt{Blanco80, Cohen81,Nikolaev00,Cioni03,Davidge03,Davidge05, Kang05,Sohn06,Jackson07, Valcheva07,Marigo08, Boyer09, Boyer19, Held10, Jung12,Lee25}), allowing exclusively for AGB star based SFHs. NIR imaging of resolved AGB stars will be easily acquired in the upcoming years with facilities such as JWST, Roman, and Euclid, providing a platform for measuring SFHs from resolved AGB stars.

In \citealt{Lee25} (hereafter \citetalias{Lee25}), we established that the SFH of M31's disk measured only from modeling AGB stars on NIR CMDs is consistent with M31's SFH measured from optical CMDs that reach below the RC from \cite{Williams17}. Though an important result on its own, it would be better to validate SFHs based on AGB stars against SFHs measured from CMDs that reach the oMSTO.  This is challenging because such a comparison needs to fulfill several criteria. First, it requires a dataset that contains enough AGB stars to measure an accurate SFH.  \citetalias{Lee25} found  that at least $\sim1000$~stars on a NIR CMD are required to measure an accurate SFH (this includes AGB stars, red helium-burning (RHeB) stars, and bright RGB). Second, the NIR CMD needs to extend $\gtrsim1$~mag below the tip of the red giant branch (TRGB). Finally, the galaxy needs to have an optical or NIR CMD that reaches below the oMSTO for validation over a similar area as the NIR data.  Most available datasets fail to simultaneously fulfill these criteria. 
The Magellanic Clouds do satisfy these criteria.  However, they were used to calibrate the \texttt{COLIBRI} AGB star models and cannot also be used for validation. 
M31 and M33 contain large numbers of AGB stars but lack oMSTO-derived SFHs or have oMSTO-based SFHs over very small areas (e.g., \citealt{Brown06, Richardson08, Benard15}). Galaxies with oMSTO-derived SFHs, such as dwarf galaxies (e.g., \citealt{Gullieuszik07a, Gullieuszik07b,Gullieuszik08b, Gullieuszik08a, Menzies08, Held10,  Savino25}) or galaxies with SFHs derived from small-area HST or JWST fields (e.g., \citealt{Dolphin03, Skillman03,deboer11, deboer12a, deboer12b, deboer14, Geha15, Munoz18}), typically lack sufficient numbers of AGB stars. 

One galaxy that approximately satisfies these criteria is WLM.  WLM is a star-forming dwarf galaxy ($M_{\star} = 4.3\times10^7 M_{\odot}$; \citealt{McConnachie12}) located at the edge of the Local Group ($d = 986$~kpc; \citealt{Lee21}). 
WLM is often considered an archetypal `isolated' dwarf galaxy (i.e., its evolution has not been shaped by significant interactions with a more massive galaxy).

WLM's SFH has been measured from the oMSTO from multiple HST and JWST fields \citep{Albers19, McQuinn24a, Cohen25}. There exists separate ground-based NIR photometry of WLM that both extends at least $\sim1$~mag below the TRGB and contains a sufficient number of AGB stars to measure an accurate SFH \citep{Lee21}. While these datasets are not perfectly matched spatially, we only need to make modest assumptions about the spatial distribution of the populations to correct for this as we detail later in this paper. We therefore undertake a comparison of the lifetime SFH of WLM derived from data that reach below the oMSTO and NIR data of only AGB stars.

This paper is organized as follows. We summarize the observations and photometry in \S \ref{sec:data}. The SFH methodology is described in \S \ref{sec:SFH}. We present our main results and compare our SFHs with those measured from deep JWST imaging in \S \ref{sec:comparison}. We present the discovery of a \added{candidate} accreted system in WLM's outer disk in \S \ref{sec:outer_burst}. We discuss radial age and metallicity gradient trends in \S \ref{sec:age_gradient}. We discuss the future prospects of using AGB stars as SFH indicators with JWST in \S \ref{sec:future}. We summarize our findings in \S \ref{sec:conclusion}.

\section{Data}\label{sec:data}

\begin{figure}
\includegraphics[width=\columnwidth]{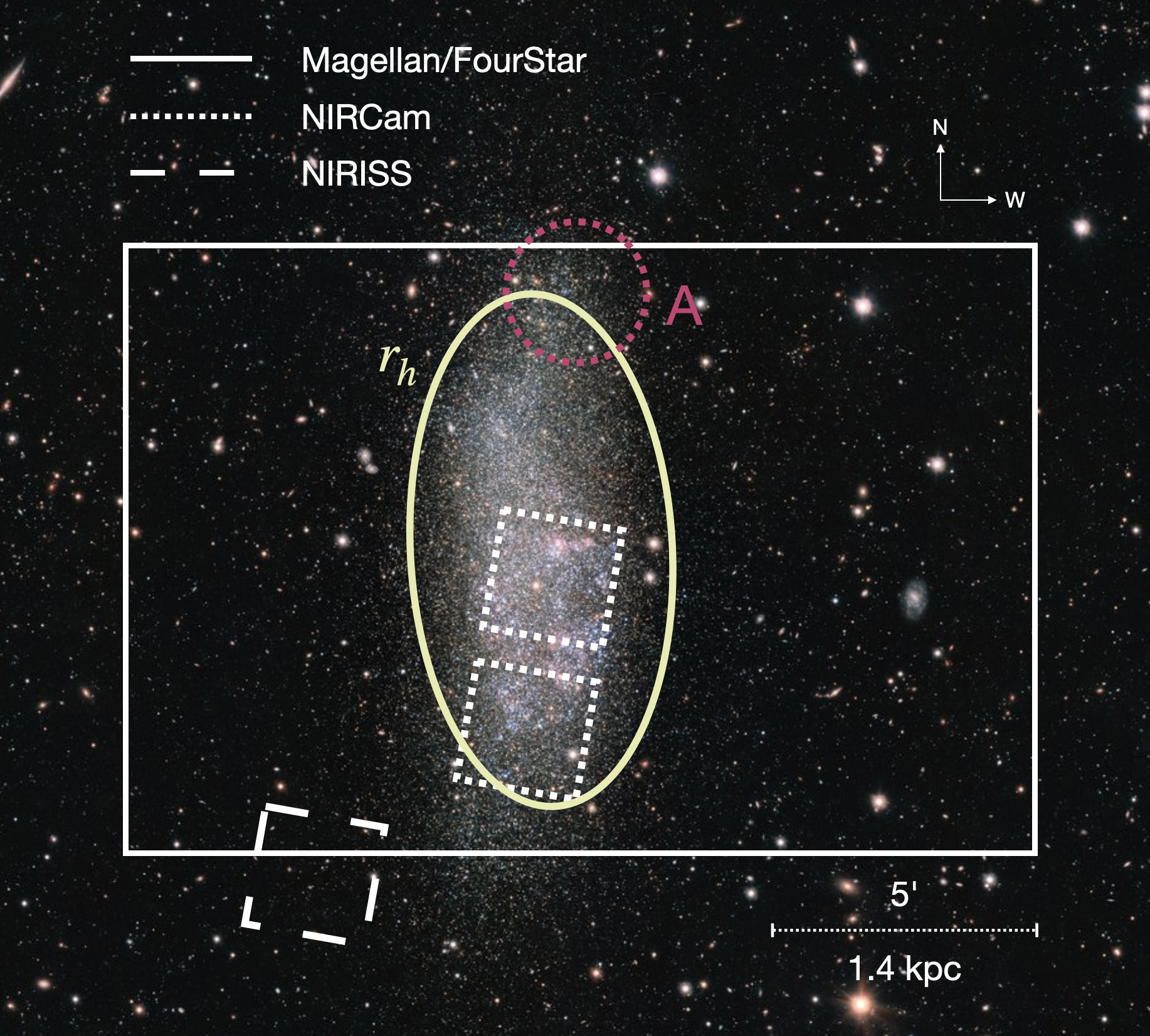}
\caption{Optical image of WLM (image credit: ESO's VST/Omegacam) overlaid with footprints of the Magellan/FourStar (solid), JWST NIRCam (dotted), and NIRISS (dashed) fields. The pale yellow line denotes WLM's half-light radius. The half-light circle of Region `A' is denoted by the pink circle and is discussed in \S \ref{sec:outer_burst}. 
}\label{fig:image}  
\end{figure}

To measure the AGB star SFH of WLM, we use  NIR ground-based imaging taken at the 6.5 m Magellan-Baade telescope at Las Campanas Observatory using the FourStar Infrared Camera \citep{Persson13} in the $J$ and $K_S$ filters. \citet{Lee21} previously used these data to measure distances to WLM using the \added{J-region asymptotic giant branch (JAGB) method, TRGB, and Cepheid period-luminosity relation}. \added{Full} details of the data acquisition and point source fitting photometric reduction can be found in that paper, but we provide a brief overview here.  \added{Three sets of observations were taken: two in 2011 with exposures of $\sim10^4$~s each and one in 2019 for calibration purposes for $\sim10$~seconds. The data were reduced using the point-spread function (PSF) photometry package DAOPHOT \citep{Stetson87} and then calibrated onto the 2MASS filter system by cross-matching the brightest stars in both catalogs. Finally, the catalog was culled of non-stellar objects based on the sharpness, error, and chi fit parameters returned by the DAOPHOT software. 
}

Figure~\ref{fig:image} shows the footprint of the FourStar field overplotted on an optical image of WLM.  This field covers WLM's entire disk and part of its stellar halo ($\sim176~\rm{arcmin}^2$/$14.5~\rm{kpc^{2}}$), and extends out to $\sim4.0r_h$ on the minor axis, \added{where $1r_h=1.3$~kpc. The half-light radius was determined by fitting ellipses to 3.6 $\mu$m Spitzer imaging, as described in \cite{Cohen25} and K. B. W. McQuinn et al. 2025 (in preparation).}

\begin{figure}
\includegraphics[width=\columnwidth]{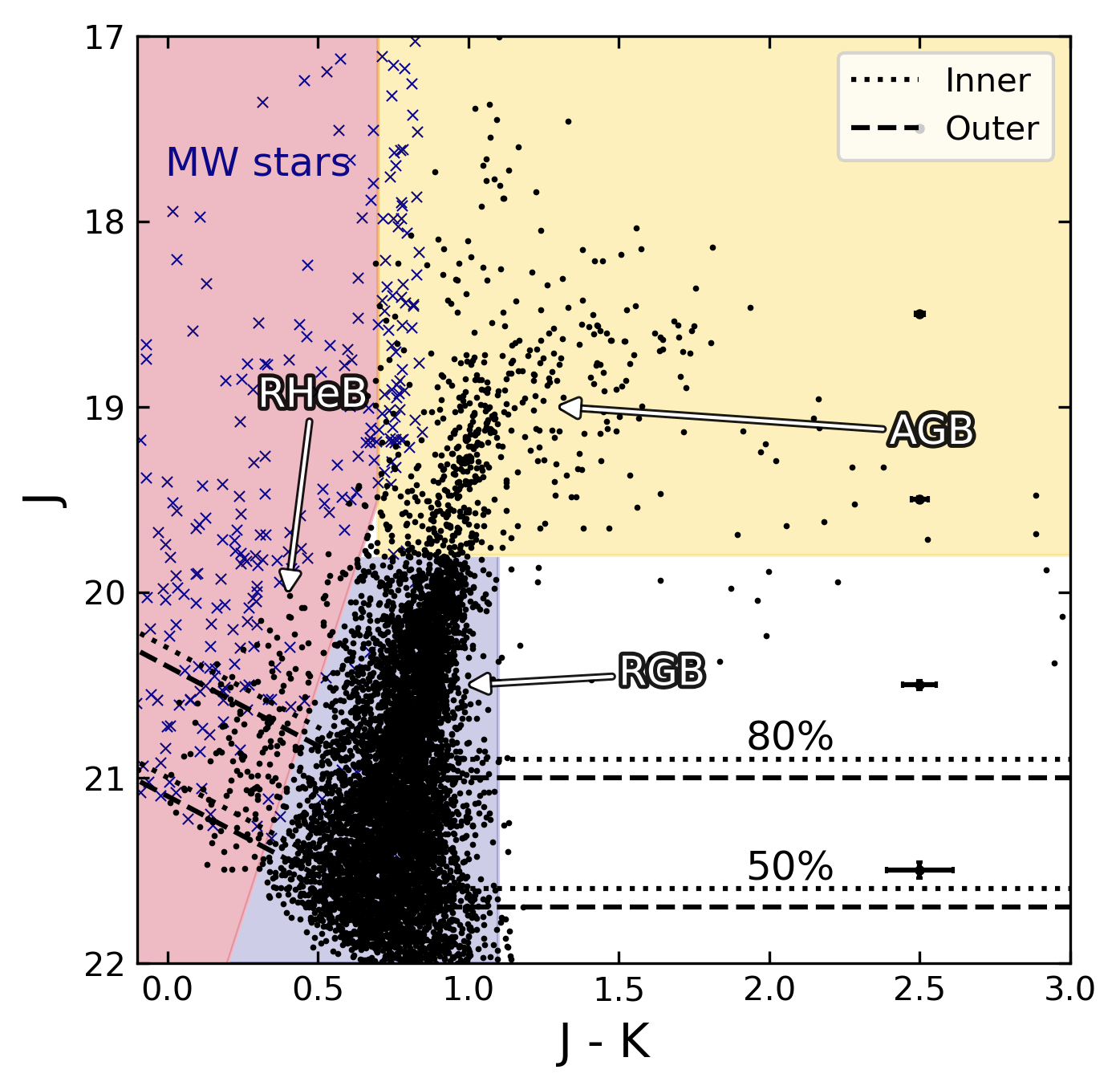}
\caption{Ground-based NIR CMD of WLM which includes $\sim13,000$~stars, of which 825 are used in our CMD modeling analysis ($J<20$~mag, $-0.2<J-K<3$~mag). Representative photometric uncertainties are shown on the right-hand side as a function of magnitude. We indicate AGB stars (yellow shading), RHeB stars (red shading), and RGB stars (purple shading). Spatial maps of these populations are shown in Figure \ref{fig:maps}. The purple x's represent stars that were matched with \textit{Gaia} and removed as foreground stars. The 80\% and 50\% completeness limits are marked with \added{dotted lines for the inner region ($r<r_h$), and dashed lines for the outer region ($r>r_h$)}. All sources have a signal-to-noise ratio (SNR) of $\ge7$ in the J band. Stars at the color and magnitude of the TRGB have an SNR $\approx85$ in the J band.
}\label{fig:cmd}  
\end{figure}

Figure~\ref{fig:cmd} shows our NIR CMD of WLM. Alongside the AGB stars (yellow shading), we indicate the locations of RGB (purple shading) and RHeB stars (red shading). The selection criteria were chosen visually, based on clear features in the CMD, such as the TRGB and the red and blue edges of the RGB.  For reference, the signal-to-noise ratio (SNR) of the photometry is $\approx85$ at the boundary between the TRGB and the faintest AGB stars ($J\sim20$).

WLM's low Galactic latitude results in a non-negligible number of Milky Way (MW) foreground stars in our NIR CMD.  We identified and removed these foreground stars using \textit{Gaia} data.  Specifically, we matched stars from our NIR CMD to sources from the \textit{Gaia} DR3 source catalog \citep{Gaia16, Gaia23} and removed matched sources closer than $1''$ and with colors bluer than $(J-K)=0.85$~mag. We selected this color cut based on a simulation of Galactic foreground stars in these filters from \texttt{TRILEGAL} \citep{Girardi05}. The removed MW  stars are shown as purple x's in Figure \ref{fig:cmd}. 

\begin{figure*}
\includegraphics[width=\textwidth]{"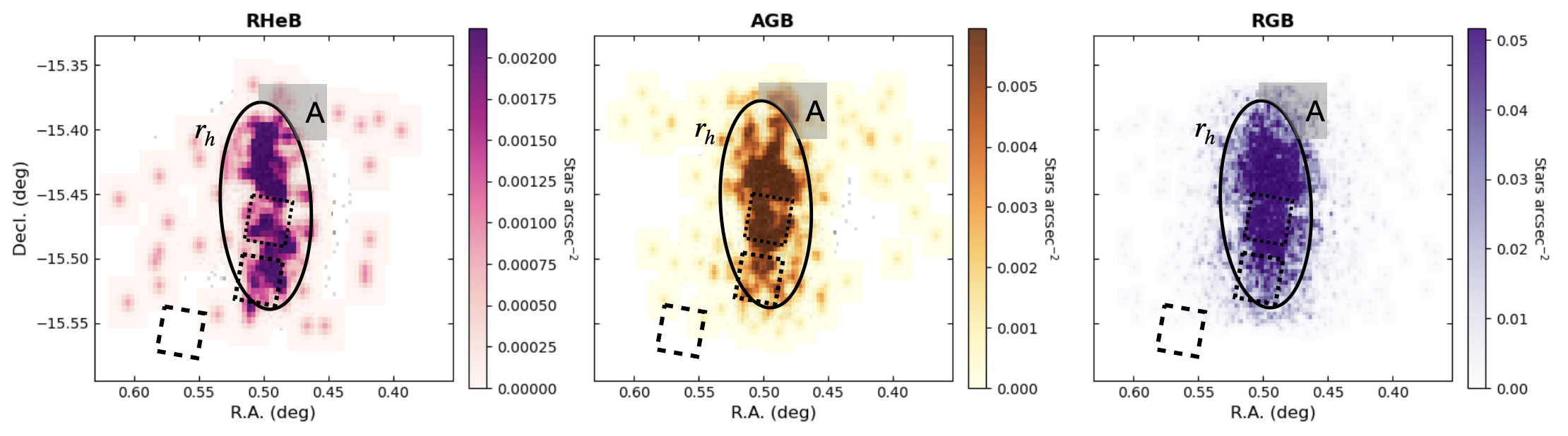"}
\caption{The spatial distribution of RHeB, AGB, and RGB stars. The photometric criteria used to select each respective stellar population is shown in Figure \ref{fig:cmd}. The NIRCam and NIRISS pointings in Figure \ref{fig:image} are shown as dotted/dashed lines, respectively. The grey highlighted region labeled ``A'' is discussed in  \S \ref{sec:outer_burst}.
}
\label{fig:maps}
\end{figure*}

To place WLM's different stellar populations (AGB, RGB, RHeB) into context, we show their spatial distributions in Figure \ref{fig:maps}.  The RGB and AGB populations have qualitatively similar spatial distributions, whereas the RHeB stars are more sparse and centrally concentrated, as is often the case in star-forming dwarf galaxies (e.g., \citealt{Bastian11}).

WLM was observed with NIRCam and NIRISS imaging as part of the JWST Early Release Science program for Resolved Stellar Populations \citep{Weisz23}. We plot footprints of the NIRCam and NIRISS fields in Figures~\ref{fig:image} and \ref{fig:maps}. 

While the JWST data reach well below the oldest MSTO, they only cover a modest fraction of WLM's total area.  The JWST data alone contain too few AGB stars for a robust SFH. Instead, we rely on the oldest MSTO-based SFHs measured from these JWST fields as a comparison point for our AGB star-based SFHs.  The JWST observations of WLM were reduced following the procedures outlined by the Early Release Science (ERS) program \citep{Weisz24}.  Extensive details of the JWST observations and photometric reductions are provided in \citet{McQuinn24a} and \citet{Cohen25}.

\section{Methodology}\label{sec:SFH}
To measure the AGB star SFH of WLM, we followed the procedure outlined in \citetalias{Lee25}, which we briefly summarize here. 

First, we fit the CMD using the CMD-modeling software package \texttt{MATCH} (Version 2.7; \citealt{Dolphin02, Dolphin12, Dolphin13, Dolphin16}). \texttt{MATCH} recovers a system's SFH by finding the linear combination of synthetic CMDs that best match the observed CMD. We generated single stellar population (SSP) models constructed from the \texttt{COLIBRI} AGB stellar isochrones (version \texttt{S\_37}; \citealt{Bressan12, Tang14, Chen14, Chen15, Marigo17, Pastorelli19, Pastorelli20}). 
These \texttt{COLIBRI} models were calibrated using observations of AGB stars in the LMC and SMC, the latter of which has a similar metallicity to WLM.   \added{These models were then folded into MATCH using one of its modes \citep[e.g.,][]{Rosenfield17} to incorporate models or CMDs that are not already internally available to it. }

We generated our SSPs using a Kroupa initial mass function \citep{Kroupa01} normalized between 0.03$M_\odot$ and 120$M_\odot$ using a star formation rate (SFR) of $0.75\rm~{M_{\odot}~yr^{-1}}$. 
\added{We probe the full lifetime SFH of WLM as simulations in Appendix B of \cite{Lee25} demonstrate that TP-AGB, E-AGB, and RGB stars can measure SFHs at lookback times of $\sim14$~Gyr ago.}
We thus generate our SSPs over an age range of $6.70<\log_{10} \rm{(t/yr)}<10.15$ with 0.1 dex resolution from  $6.7<\log_{10} \rm{(t/yr)}<9.0$ and 0.05 dex resolution from $9.00<\log_{10} \rm{(t/yr)<10.15}.$ 
These SSPs spanned metallicities from $-2.0\le \rm{[M/H]}\le-0.5$ with a resolution of 0.1~dex. Following common practice in the literature (e.g., \citealt{Weisz14b,Savino23,McQuinn24a, Cohen25}), we required the metallicity to increase monotonically as a function of time (with a spread of 0.1~dex at each age), with a metallicity range of $-2.0\le \rm{[M/H]}\le-1.4$ at the oldest time to a present-day metallicity range of $-1.0\le \rm{[M/H]}\le-0.5$. 

For the purposes of validation, we chose these metallicity prior ranges to match those of \citealt{Cohen25} (hereafter \citetalias{Cohen25}), who measured WLM's SFH from the JWST NIRCam and NIRISS data.  We did not fix the metallicities or age-metallicity relations (AMRs) to be identical, only the starting and ending ranges.  These values are commonly adopted in other LG SFH determinations \citep[e.g.,][]{Monelli10a, Monelli16, Weisz14b, Skillman17,Hargis20, Savino23, Savino25, McQuinn24b,McQuinn24a, Cohen25}.

We modeled observational effects in the CMD (photometric errors and incompleteness) using $3 \times 10^4$ artificial star tests (ASTs). The ASTs were distributed uniformly both in color and magnitude over the CMD and spatially throughout the Magellan/FourStar footprint.  We inserted only 5\% of the total stars per image to avoid effects of self-crowding, \added{as is commonly done in the literature (e.g., \citealt{Harris08, Piatti12}).} 
An artificial star was only considered recovered if it passed all the photometric quality cuts applied to the photometry. More details on the photometric quality cuts are available in \citet{Lee21}.

We adopted a WLM distance modulus of $\mu_0=24.93$~mag ($968$~kpc).  This is the same value used by \citetalias{Cohen25}. This choice provides for self-consistency in SFH comparisons. This distance was derived from the TRGB \citep{Albers19}, and is consistent with commonly adopted distances to WLM measured with the TRGB, JAGB method, and Cepheids (e.g., \citealt{McConnachie05, Jacobs09,  Lee21, Anand21}).

We adopted a Galactic extinction value of $A_V=0.12$~mag from \citet{Schlafly11}. For a \cite{Cardelli89} extinction law and an $R_V=3.1$, this corresponds to $A_J=0.03$~mag and $A_K=0.01$~mag. We applied these corrections directly to our photometry before modeling the CMD.  We did not solve for \added{internal} extinction as part of the CMD modeling. 
\added{Tests with MATCH by \cite{Albers19, McQuinn24a, Cohen25} found zero internal extinction in HST and JWST fields of WLM out to   $\sim3r_h$. Furthermore, \cite{Lee21} measured the TRGB in three non-overlapping regions within the same FourStar dataset analyzed in this paper, finding non-significant differences in their measured values, and thus concluding that internal reddening in WLM is negligible. } Therefore, we did not include reddening internal to WLM in our modeling as these studies have already demonstrated that WLM has no measurable internal dust. \added{Finally, any self-extinction due to the circumstellar dust in the extended envelopes of AGB stars are already directly built into the \texttt{COLIBRI} isochrones \citep{Marigo08,Marigo17, Pastorelli19, Pastorelli20}.}

We converted the CMD to a Hess diagram with bin sizes of $0.05 \times 0.05$~mag in color and magnitude, respectively.  \added{The bin sizes were chosen based on the high SNR of our data (SNR $>85$).}
We then fit the binned CMD down to the 90\% completeness limit of $J=20.0$~mag. 
Our fits also included a CMD model of \added{$\sim3300$} Galactic foreground stars generated from the \texttt{TRILEGAL} simulation \citep{Girardi05}.  
\added{While we already removed some foreground contaminants by cross-matching with \textit{Gaia} as described in \S \ref{sec:data}, \cite{Barmby23} showed that \textit{Gaia} foreground removal is incomplete for magnitudes around than the TRGB for galaxies at distances $>450$~kpc. Thus, we chose to include this additional foreground model in our CMD modeling to account for any additional faint foreground contaminants. }

To accurately account for spatially-dependent observational biases and completeness, we measured the SFH of the inner region of WLM ($r<r_h$) and outer region of WLM ($r>r_h$) separately, and then combined their results to determine the total SFH. In total, 825 stars were used for the SFH fits in the inner and outer regions combined, including \added{thermally-pulsating AGB (TP-AGB) early AGB (E-AGB), RGB, and RHeB stars. Most of the age SFH information is derived from TP-AGB stars, except for the very oldest ages ($>10$~Gyr) for which the E-AGB and RGB stars provide leverage.}
\added{We note E-AGB stars are expected to be $\sim2-20\times$ more abundant than TP-AGB stars \citep{ Vassiliadis93, Habing03}.
}

\begin{figure}
\includegraphics[width=\columnwidth]{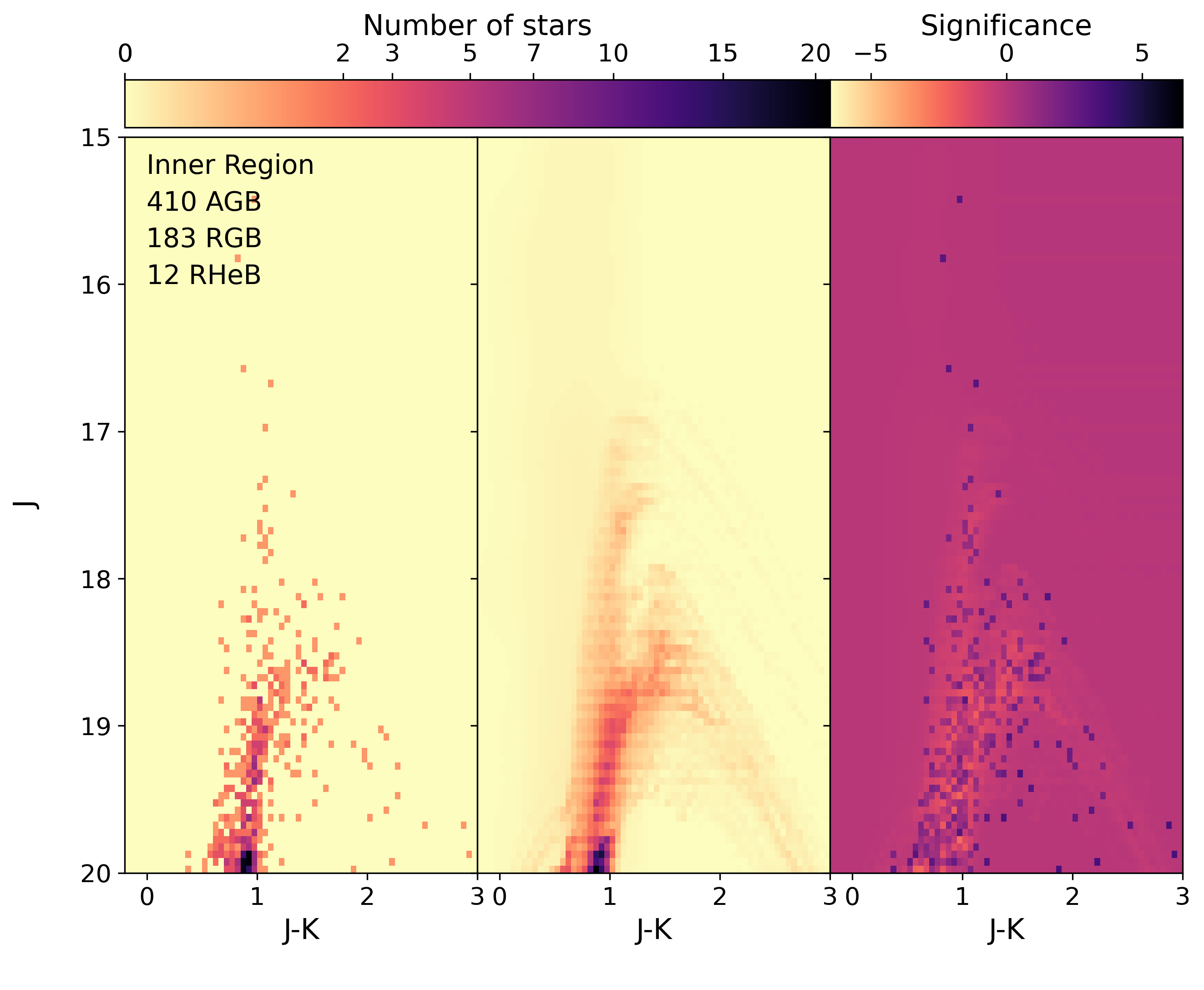}
\includegraphics[width=\columnwidth]{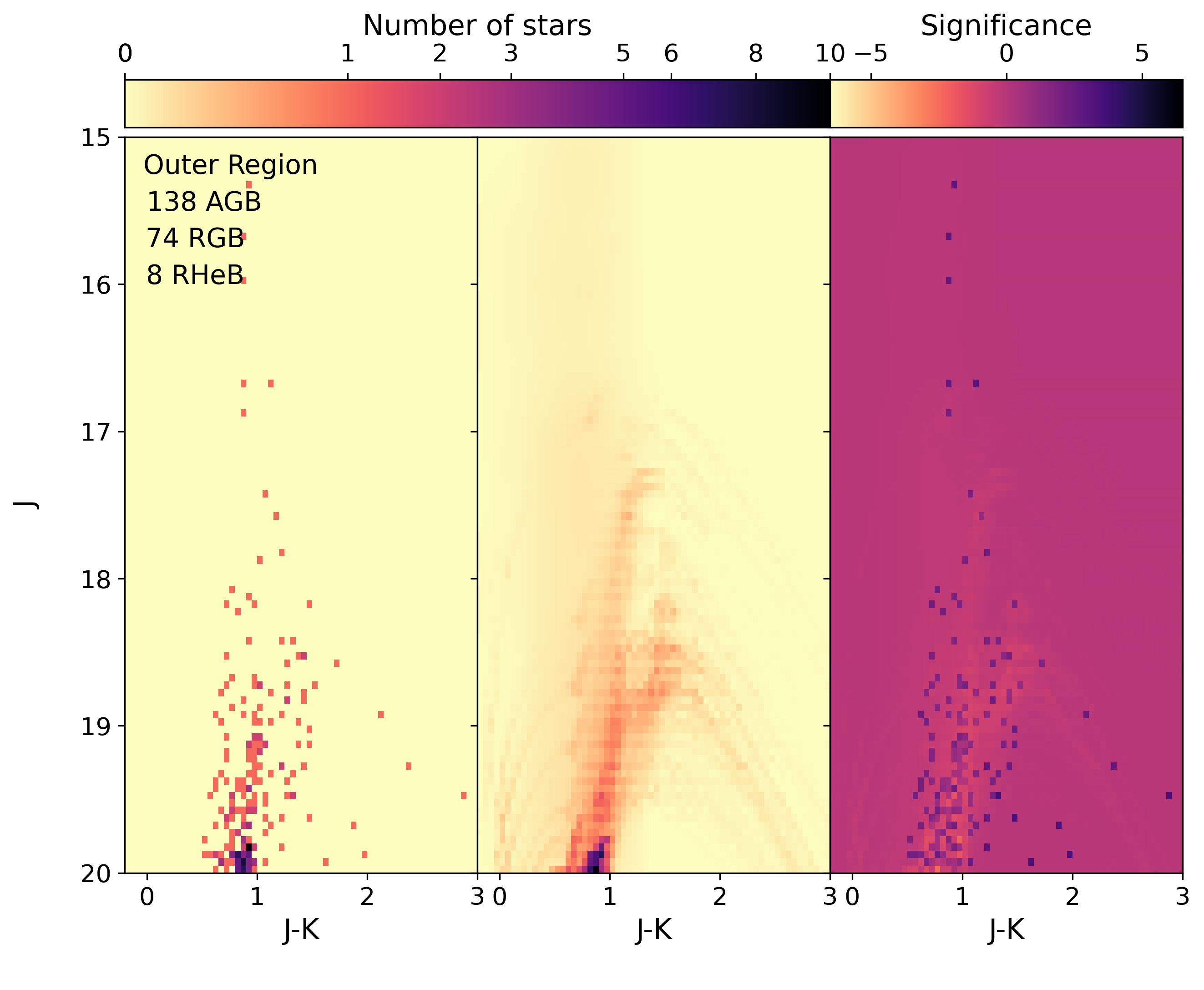}
\caption{Hess diagrams for our AGB star SFH fits for the observed data (left panel), best-fit linear combination of SSPs (middle panel), and residual CMDs (right panel) for the inner region ($r<r_h$, top) and outer region ($r>r_h$, bottom). The residuals are expressed in Poisson standard deviations. \added{The number of AGB, RGB, and RHeB stars in each CMD is also listed in the top lefthand corner of each plot. }
}\label{fig:fit_cmd}  
\end{figure}

Results of the CMD fits for the inner and outer regions are shown in Figure \ref{fig:fit_cmd}. There is good overall agreement between the  observed and model CMDs in both cases.  The residual significance plots (right panels) show no notable systematic structures and there appear to be no areas of either the inner or outer CMDs that are poorly modeled.

We calculated random uncertainties using the hybrid Monte Carlo process described in \citet{Dolphin13}. 
As discussed in \citetalias{Lee25}, systematic uncertainties for SFHs measured with \texttt{MATCH} are usually computed following the procedure outlined in \citet{Dolphin13}.  However, because there are no other sufficiently realistic AGB star models available, we were unable to adopt this approach and we only consider random uncertainties in this paper.

\section{The Star formation history of WLM}\label{sec:comparison}

\subsection{The Global SFH of WLM from AGB Stars}\label{subsec:agb_sfh}

\begin{figure}
\includegraphics[width=\columnwidth]{"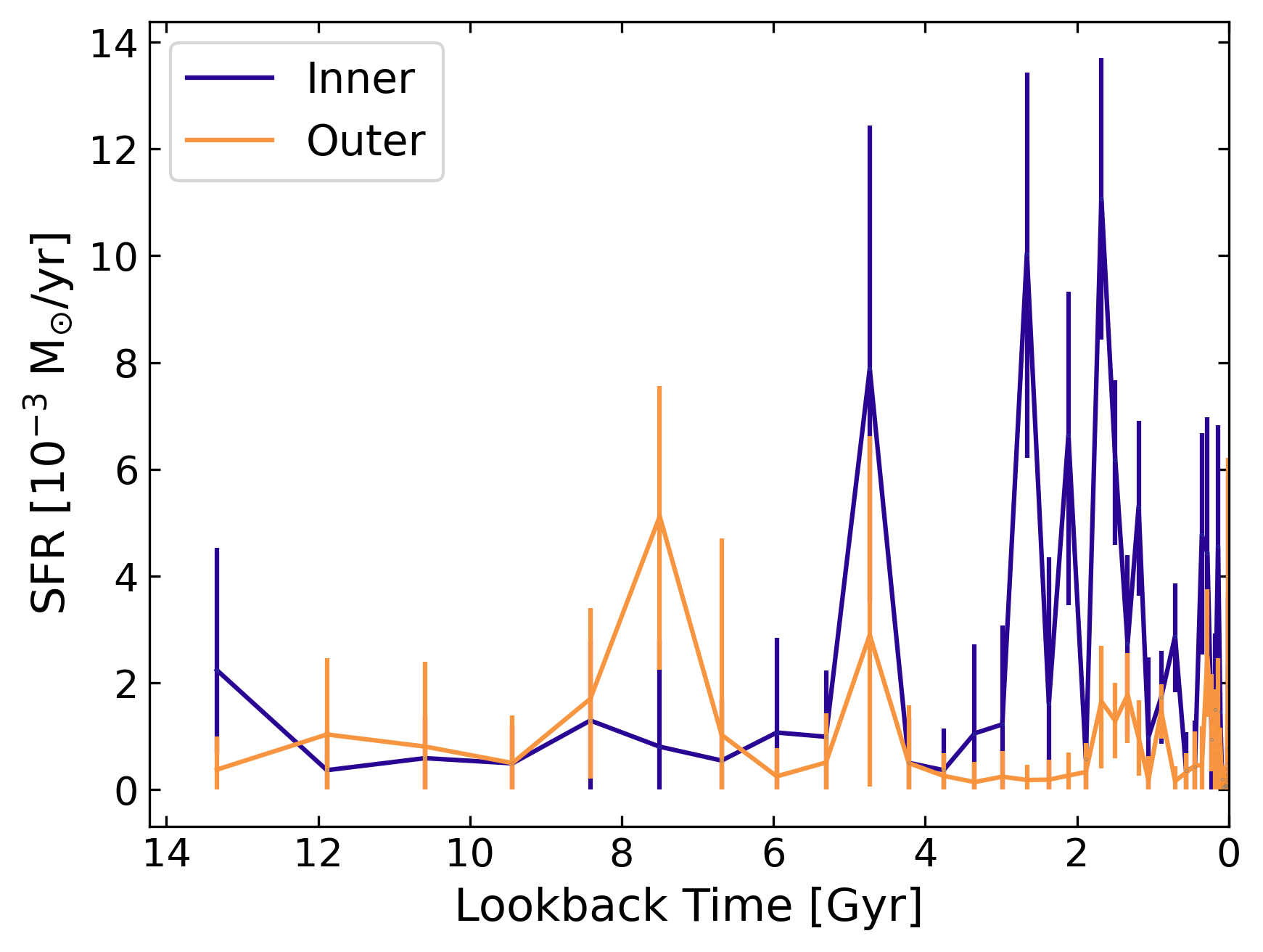"}
\includegraphics[width=\columnwidth]{"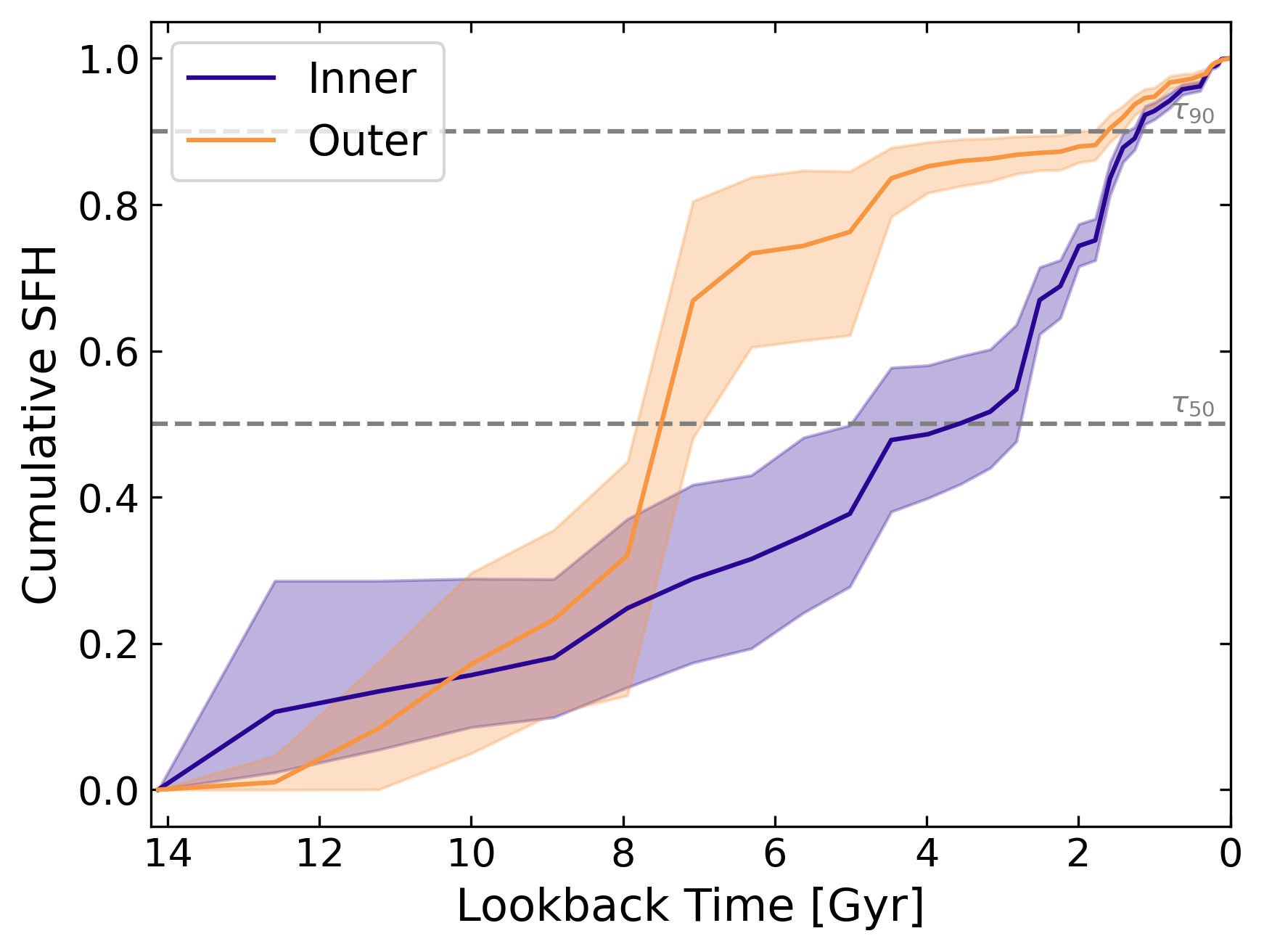"}
\includegraphics[width=\columnwidth]{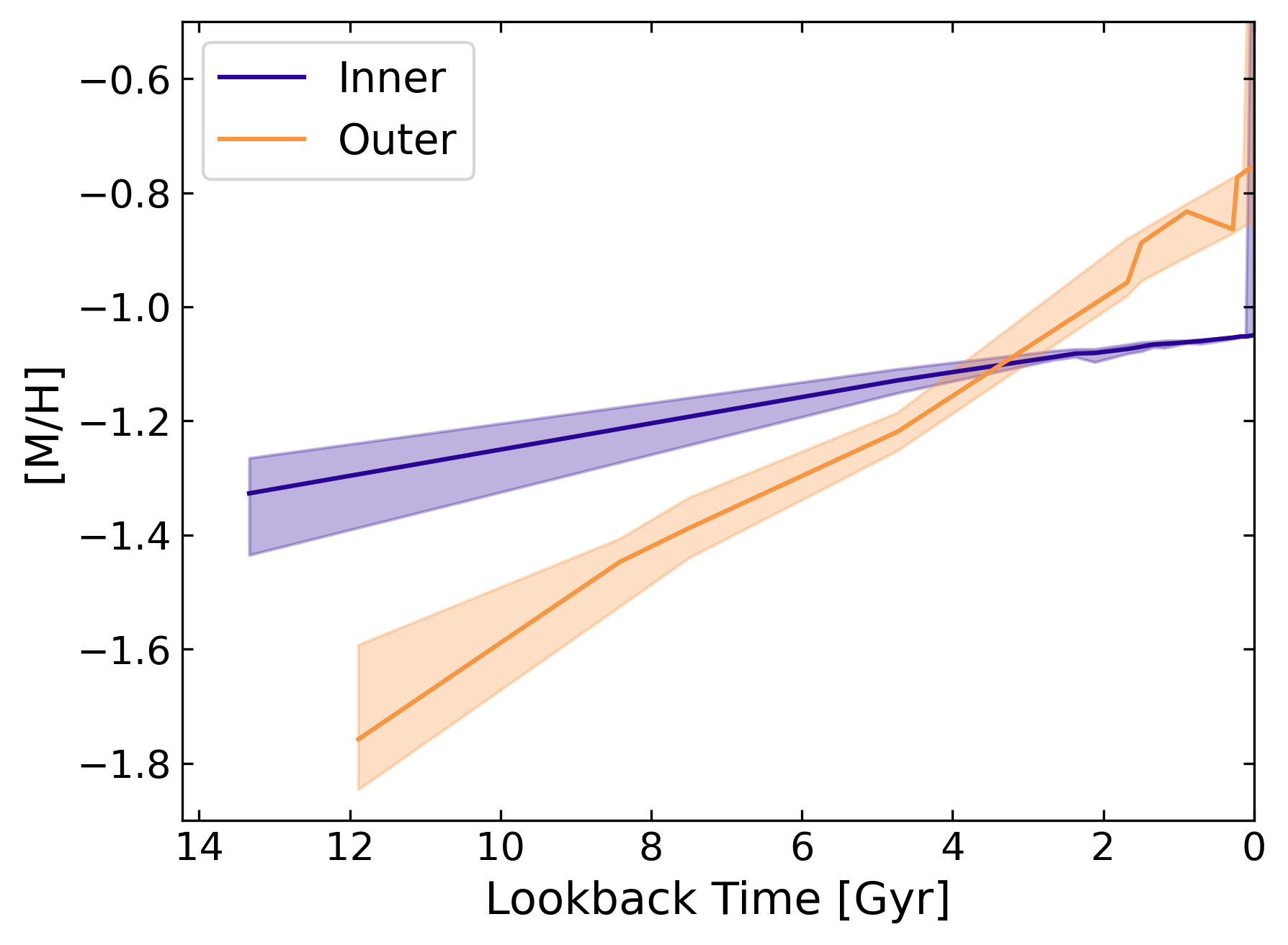}
\caption{(Top) SFR as a function of time. (Middle) Cumulative SFH. \added{$\tau_{90}$ and $\tau_{50}$ are plotted as dashed grey lines.} (Bottom) AMR \added{where only the metallicity points in which the  SFR is
greater than $\gtrsim2\%$ of the total star formation are plotted.} All plots show fits derived from the AGB stars for the inner ($r<r_h$) and outer regions ($r>r_h$) of WLM. } \label{fig:outer_inner} 
\end{figure}

 The star formation rates, cumulative star formation history, and age-metallicity relation for the outer and inner regions are shown in Figure \ref{fig:outer_inner}.  The plotted random uncertainties correspond to the 68\% confidence interval. 
 We list measured values of $\tau_{50}$ and $\tau_{90}$ and their uncertainties for the inner region, outer region, and combined result in Table \ref{tab:SFR_frac}. We discuss further the age gradient in WLM in \S \ref{sec:age_gradient}.  We note the outer region shows a burst of star formation $\sim8$~Gyr ago, not seen in the inner region. We further discuss the potential origin of this burst in  \S \ref{sec:outer_burst}.

\begin{figure}
\includegraphics[width=\columnwidth]{"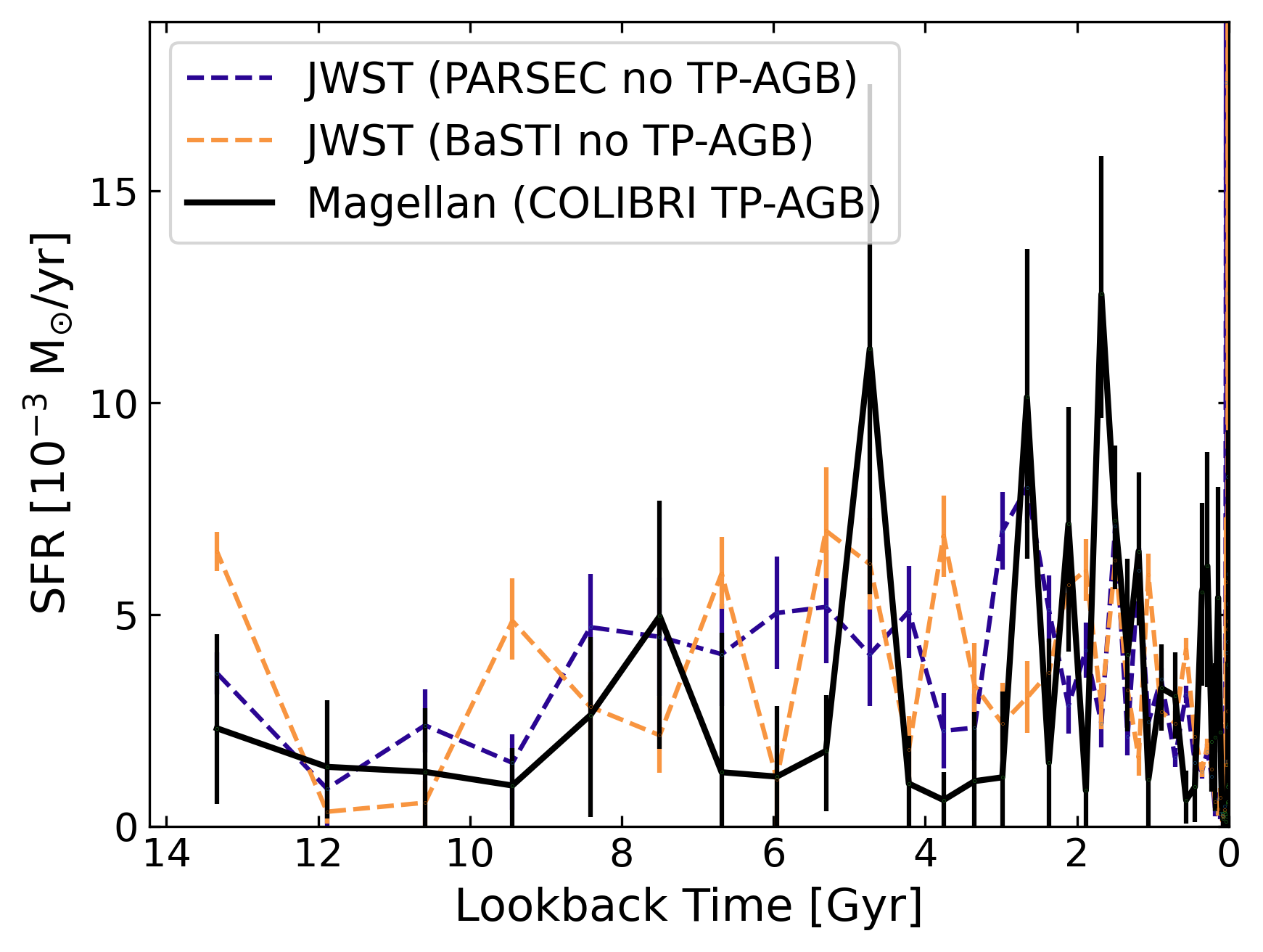"}
\includegraphics[width=\columnwidth]{"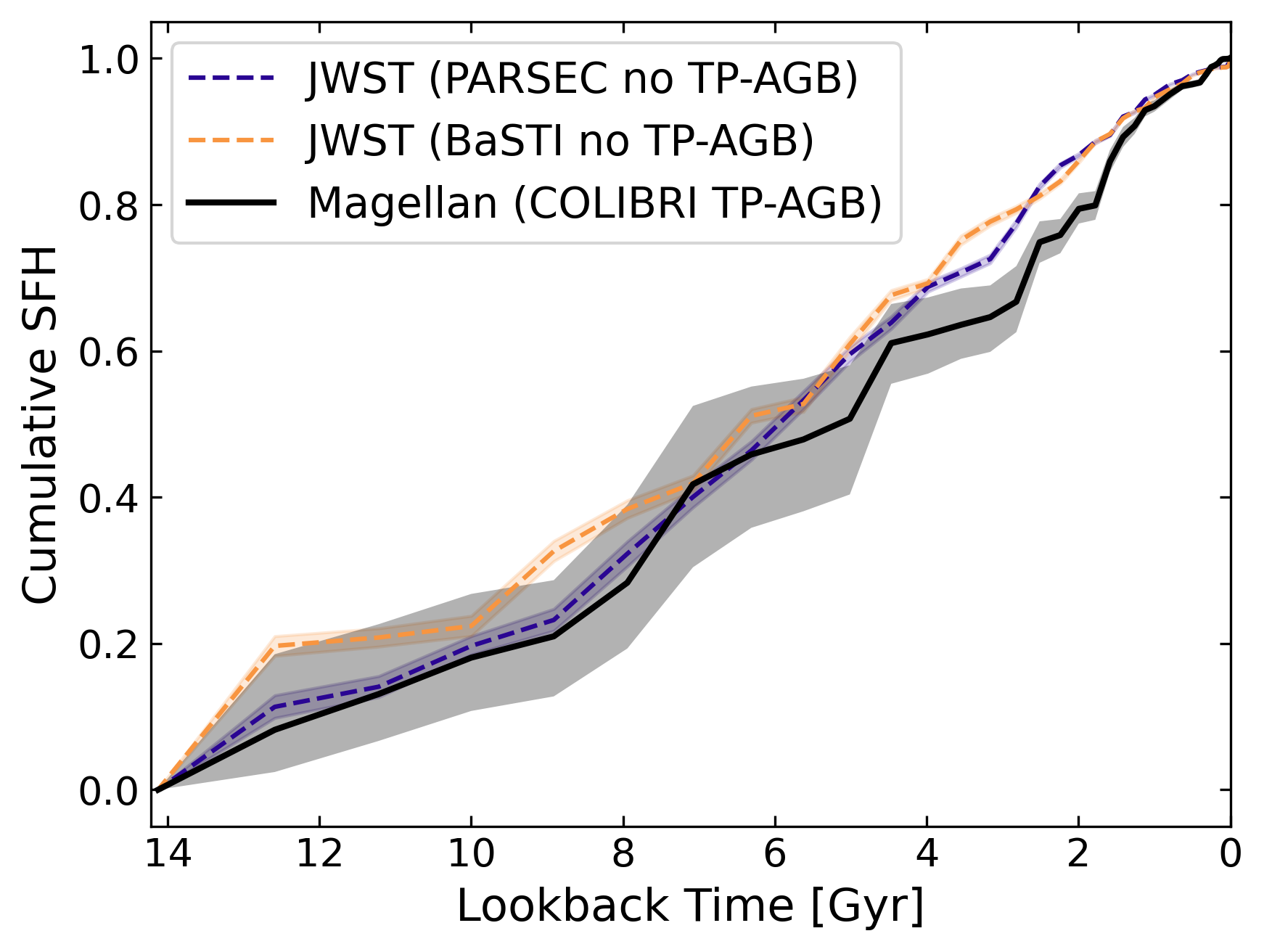"}
\includegraphics[width=\columnwidth]{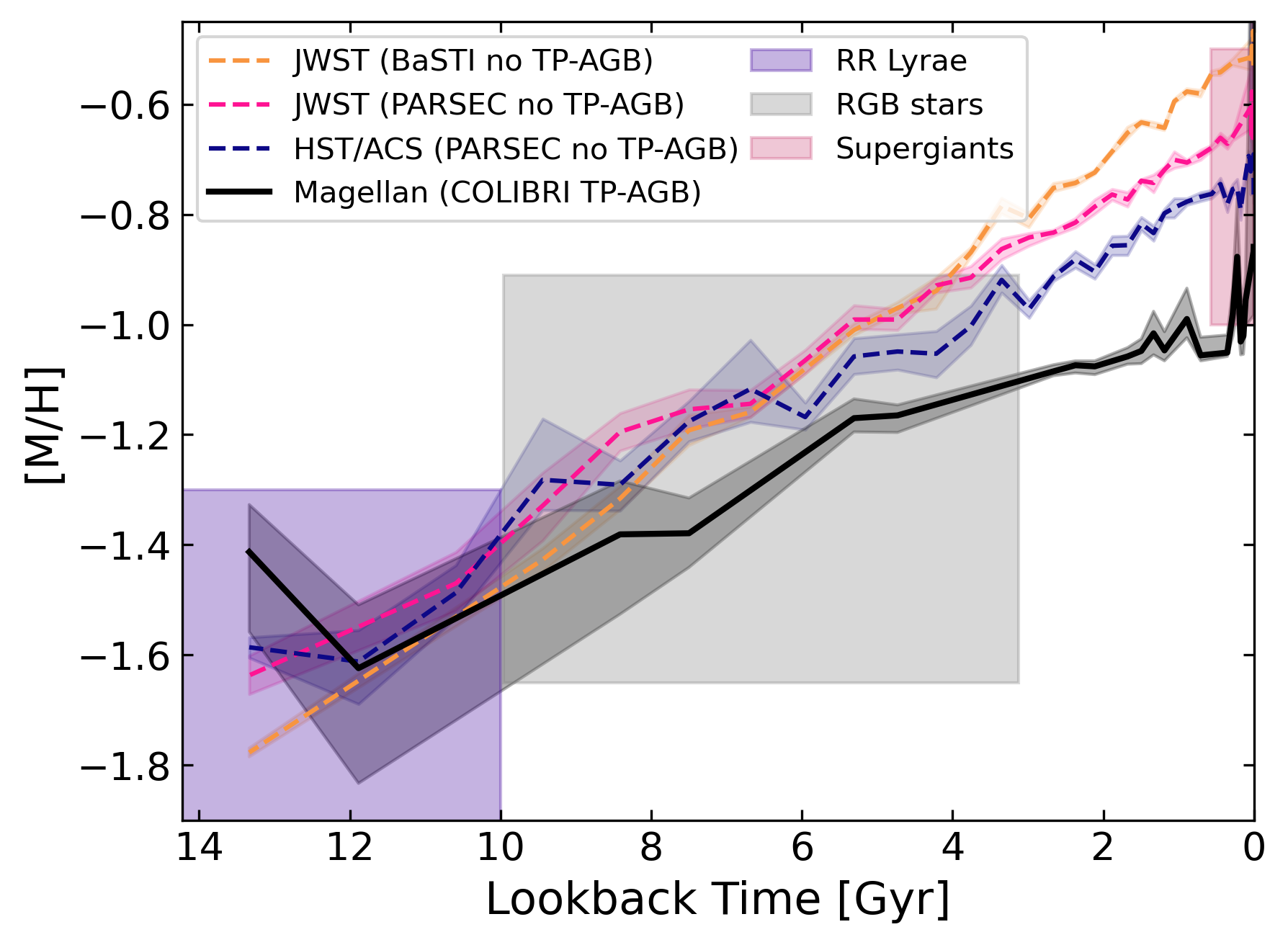}
\caption{Top panel: SFR as a function of time. Middle panel: global cumulative SFH. Bottom panel: AMR derived from the AGB stars compared with the oMSTO SFHs. Spectroscopically determined metallicity ranges from  RR Lyrae \citep{Sarajedini23}, RGB stars \citep{Leaman09}, and supergiants \citep{Urbaneja08} are shown for reference.  Age ranges for the spectroscopic measurements are taken from each of the indicated papers. The magenta and orange lines represent fits derived from the oMSTO using the PARSEC and BaSTI models, respectively, from JWST NIRCam and NIRISS imaging analyzed in \citetalias{Cohen25}, that have been scaled to match the field of view (FOV) of our ground-based observations. The indigo line represents the fit derived from the oMSTO using the PARSEC models from HST/ACS imaging analyzed in \citetalias{Cohen25}.} \label{fig:sfh} 
\end{figure}

\begin{deluxetable}{ccc}
\tablecaption{Values of $\tau_{50}$ and $\tau_{90}$ Obtained from the AGB star SFH and oMSTO fits}\label{tab:SFR_frac}
\tablehead{
\colhead{Model} & 
\colhead{$\tau_{50}$ (Gyr)} &
\colhead{$\tau_{90}$ (Gyr)}}
\startdata
AGB (Global) & $5.16_{-0.50}^{+2.07}$ & $1.33_{-0.09}^{+0.11}$\\
PARSEC Full CMD & $5.94_{-0.31}^{+0.37}$ & $1.55_{-0.13}^{+0.03}$\\
BaSTI Full CMD & $6.40_{-0.09}^{+0.68}$ & $1.56_{-0.14}^{+0.03}$\\
\hline
AGB (Inner)  & $3.58_{-0.81}^{+1.42}$ & $1.22_{-0.09}^{+0.14}$\\
PARSEC NIRCam Full CMD & $4.87_{-0.40}^{+0.14}$ & $1.25_{-0.12}^{+0.01}$\\
BaSTI NIRCam Full CMD  & $4.99_{-0.52}^{+0.06}$ & $1.24_{-0.12}^{+0.02}$\\
\hline
AGB (Outer w/ Region A) & $7.48_{-0.53}^{+0.46}$ & $1.62_{-0.19}^{+0.40}$\\
AGB (Outer w/o Region A) & $7.91_{-1.12}^{+1.65}$ & $1.70_{-0.12}^{+2.31}$\\
PARSEC NIRISS Full CMD & $7.76_{-0.68}^{+0.31}$ & $2.90_{-0.09}^{+0.27}$\\
BaSTI NIRISS Full CMD & $8.87_{-0.92}^{+0.23}$ & $3.11_{-0.29}^{+0.20}$
\enddata
\tablerefs{PARSEC and BaSTI values are from \cite{Cohen25}.}
\end{deluxetable}

Figure \ref{fig:sfh} presents the best-fit global cumulative AGB star SFH, SFR as a function of time, and the AMR (thick black line), for the inner and outer regions combined. 
From the cumulative SFH, we see that WLM had low level star formation until a burst $\sim8$~Gyr ago, followed by larger bursts $\sim5, 3$, and 1.5~Gyr ago. 
We recovered a total stellar mass in the full FOV of $M_{*}=3.9\pm0.3\times 10^7 M_{\odot}$. We converted this to a present-day mass by correcting for stellar evolution mass loss, which \cite{Vincenzo16} estimated to be $\sim41$\% for low-metallicity galaxies. This calculation results in a present-day mass of $M_{*,z=0}=2.3\pm0.2\times10^7~M_{\odot}$. 
This value agrees well with the stellar mass of WLM of  $2.8\pm0.8\times10^7~M_{\odot}$ calculated from WLM's 3.6 $\mu$m luminosity in \citealt{Cook14}, which we standardized with the same distance modulus used in this study. 

The AMR in Figure \ref{fig:sfh} shows the overall metallicity evolution.
\added{We only plotted metallicity data points  if the star formation at that given age exceeded $\gtrsim2\%$ of the total star formation, as is customary in the literature (e.g., \citealt{McQuinn15, Albers19, Rathi20}).}
The early populations of WLM were metal-poor at $[M/H]\approx-1.5$~dex. When WLM had formed 50\% of its mass, the galaxy had enriched to a metallicity of $[M/H]\approx-1.2$~dex. Finally, we found the present-day global metallicity of WLM to be $[M/H]\approx-0.9$~dex. We compare these values to metallicity estimates from the literature in  \S \ref{subsec:compare_lit}.

\subsection{Comparison to the JWST-based SFH}\label{subsec:compare_lit}

We now consider how the AGB star SFH compares to the SFH of WLM derived from the oldest MSTO. To validate our results against the SFH measured from the oMSTO in \citetalias{Cohen25}, we first needed to scale their SFR to match the stellar mass covered in our FOV. As shown in Figure \ref{fig:image}, the \citetalias{Cohen25} SFH is based on a NIRCam field which covers a portion of WLM's disk and a NIRISS field which lies $\sim1.5$~kpc to the southeast of WLM's central region. 

For the purposes of this exercise, we adopted the NIRCam-based SFH as the SFH for all area within $\lesssim1r_h$ and multiplied it by a factor of $\sim2.0$ (i.e., \added{the total mass derived in the inner region divided by the mass derived in the NIRCam PARSEC field}). Similarly, we assumed that the star formation of the NIRISS field represents the star formation outside of $1r_h$ and multiplied it by \added{$\sim71.1$} (i.e., \added{the total mass derived in the outer region divided by the mass derived in the NIRISS PARSEC field}). These are clearly approximations, particularly given some of the azimuthal and radial gradients reported by \citetalias{Cohen25}.  
Spatially-matched data that include AGB stars and the oldest MSTO would be ideal (e.g., with Roman), but as discussed in \S \ref{sec:intro}, such datasets are currently not available.

In addition to the results from the AGB star SFH as discussed in \S \ref{subsec:agb_sfh}, Figure \ref{fig:sfh} shows these best-fit results based on the oMSTO from \citetalias{Cohen25} derived from both the BaSTI \citep{Hidalgo18} and PARSEC \citep{Bressan12} stellar evolution libraries. We plot the 68\% random uncertainties reported by \citetalias{Cohen25}.  The \citetalias{Cohen25} SFHs and uncertainties have been scaled up by the \added{mass factors described in the previous paragraph.}

The AGB star SFH and oMSTO SFHs from both the PARSEC and BaSTI models are in excellent agreement. Like the oMSTO-derived SFHs, the AGB star SFH recovers that WLM formed $20\%$ of its stellar mass within the first $\sim4$~Gyr of its history, and then $\sim50\%$ of its stellar mass within the last $\sim5-6$~Gyr.  There are minor differences from $\sim 1-4$~Gyr ago, though this is the only time at which the SFHs disagree beyond the 68\% uncertainties on the AGB star SFHs.  We do not undertake a detailed comparison of recent star formation, as our approach to extrapolating the JWST fields will be far less accurate for younger populations that are not well-mixed.  Importantly, the AGB star SFH is well within the model-to-model variations from oMSTO CMD analysis, which is what \citetalias{Cohen25} use to gauge systematic uncertainties in the oldest MSTO SFHs. 

\added{We find good overall agreement between the AGB star SFR and oMSTO SFRs as a function of time, with a few minor differences. For example, the timing and amplitude of the AGB star SFR bursts  $\sim13-14$~Gyr ago, $\sim7-8$~Gyr ago and $\sim5$~Gyr ago are slightly different. Such differences in the exact timing of star formation from CMD modeling are often observed when using different stellar models or CMD fitting codes (e.g., \citealt{Weisz11, Dolphin12, Cole14, Skillman17, Savino23}). As a demonstration of this, any differences in the timing and amplitude of SF bursts between the AGB star SFRs and either oMSTO-based SFRs are equal or smaller than the differences between the BaSTI and PARSEC oMSTO-based SFRs. }

In terms of SFH summary statistics (see Table \ref{tab:SFR_frac}), all three fits (oldest MSTO using BaSTI and PARSEC, AGB star) yield consistent values of $\tau_{50}\approx 5-6$~Gyr ago and $\tau_{90}\approx1.3-1.6$~Gyr ago.  Uncertainties on the AGB star-based values are larger for $\tau_{50}$ than those from the oldest MSTO.  This is likely the result of shot noise and perhaps some age-metallicity degeneracy effects as only 825 stars are being used to measure the AGB star SFH. Uncertainties on $\tau_{90}$ are comparable between all three SFHs.

All three AMRs show the same qualitative trends (i.e., steady enrichment as a function of linear lookback time), though with modest offsets from one another. Compared to the AGB star based metallicity, the oldest MSTO-based AMRs are slightly more metal-poor at the very oldest ages and show steeper enrichment.  To place all three AMRs into context, we overplot several external stellar metallicity estimates for WLM.   At ancient times, we include metallicities derived from the periods of 90 RR Lyrae variables, which ranged from $-2.4\lesssim\rm{[M/H]\lesssim-1.3}$ with a mean metallicity of $\rm{[Fe/H]}=-1.74$ \citep{Sarajedini23}.  We assume RR Lyrae corresponds to stars older than $\sim10$~Gyr.  For intermediate ages, we include RGB star metallicities derived from spectroscopy of WLM red giants from  \cite{Leaman09}, who derived a mean metallicity of $\rm{[Fe/H]}=-1.28$ ($\sigma=0.37$).  We plot these over the age range provided in the \cite{Leaman09} paper.  Finally, we include present day stellar metallicity estimates from \cite{Urbaneja08}.  They derived a stellar metallicity range from six BA-type supergiants of $-1.0\lesssim\rm{[M/H]}\lesssim-0.5$ with a mean stellar metallicity of $\rm{[M/H]}=-0.87$.

All three CMD-based AMRs are within the ranges reported by these independent metallicity estimates.  
The modest offsets and slightly different AMRs between the AGB stars and oldest MSTO methods are interesting, but without more suitable datasets \added{(i.e., see the criteria listed in \S \ref{sec:intro})} it is challenging to explore the causes of any differences (e.g., \textit{does the small number of stars play a role? How does dredge up affect AGB star metallicities?}).
For the present purposes, the AMRs from the three approaches are in good agreement, and certainly are comparable to, or better than, AMRs measured historically from different CMD fitting codes. \added{For example, \cite{Skillman03,  Garling25} showed AMRs produced by independent CMD-fitting codes can differ by $\lesssim0.4$~dex, which is comparable to the differences seen between the AMRs shown in Figure \ref{fig:sfh}. } 
\added{Furthermore, there may also be minor differences in the derived AMRs arising from the different fields of view for each field, although this effect should be small as WLM is known to have a weak metallicity gradient \citep{Leaman13}.}
We also include the AMR derived from the oMSTO using the PARSEC model from HST/ACS imaging analyzed in \citetalias{Cohen25} as an additional comparison point, demonstrating that differences between the AMRs derived from HST and JWST are comparable to differences between the AMRs derived from HST and the AGB stars.

\section{An Accretion Event in WLM's Northwestern Outer Disk?}\label{sec:outer_burst}

While inspecting the SFHs of the inner and outer fields shown in Figure \ref{fig:outer_inner}, we noticed a burst of star formation at intermediate ages ($\sim8$~Gyr) recovered only in the outer field. This burst was not seen in the outer fields of the \citetalias{Cohen25} analysis, who analyzed HST WFC3/UVIS and JWST NIRISS fields in the southeast and western outer regions of WLM, both at $R\approx2.5r_h$. Nor was it observed in the central HST/ACS field analyzed by \cite{Albers19}.

\begin{figure}
\includegraphics[width=\columnwidth]{"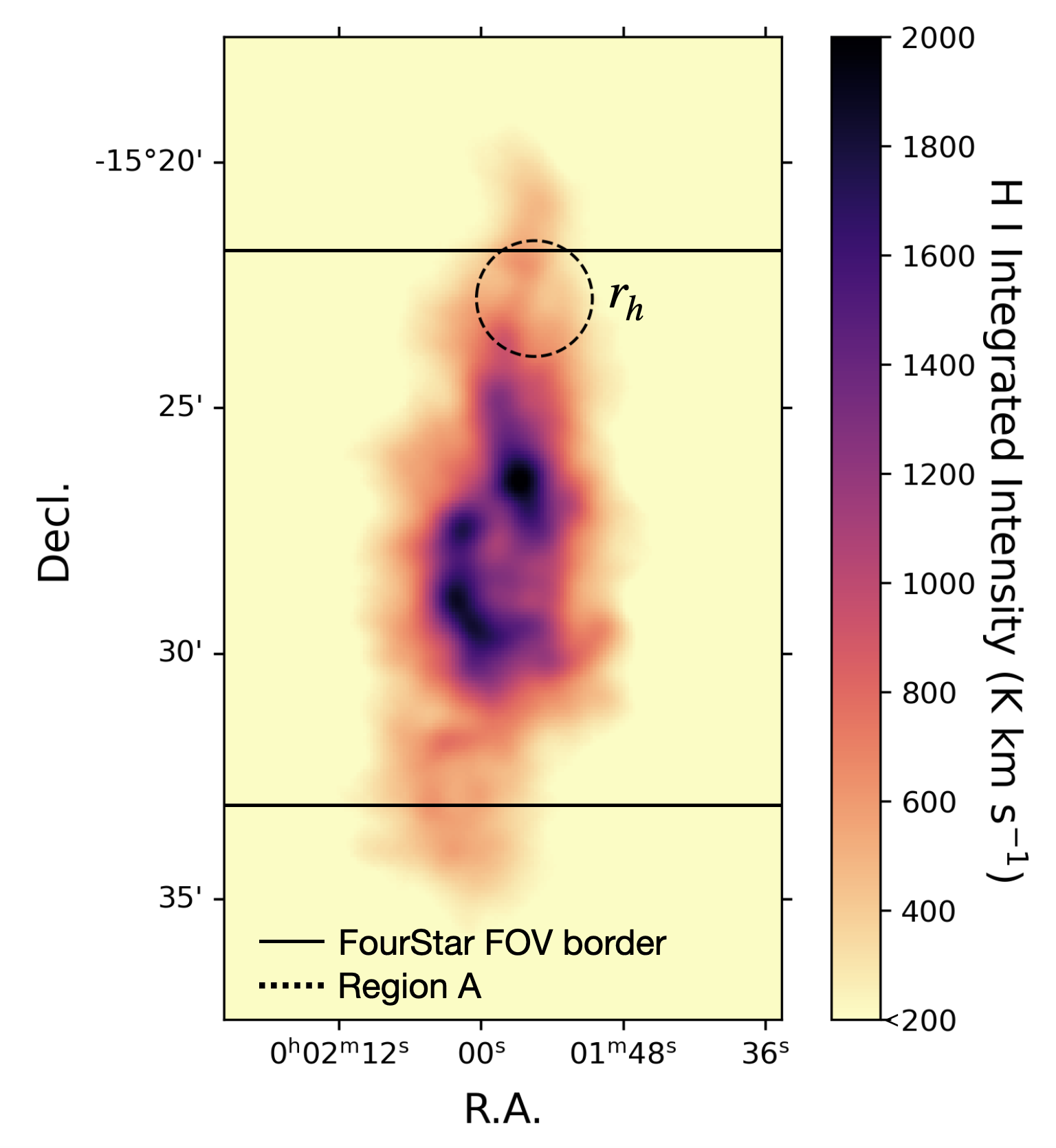"}
\caption{H~{\sc i} integrated intensity map from \cite{Koch25} overlaid with the Magellan/FourStar footprint (solid lines). The half-light circle of Region `A' is denoted by the dotted line. } \label{fig:h1} 
\end{figure}

By examining the CMD and spatial distribution of stars from our NIR data, we found that this burst appears to be due to an over-density of 49 AGB stars located in the northwestern outer disk of WLM, which we label region `A' in Figure \ref{fig:maps}.  It is visible in optical images, such as in Figure \ref{fig:image} (see also \citealt{Jackson07}). Region A also spatially coincides with a known warp in H~{\sc i} observations (e.g., \citealt{Jackson04, Koch25}), as shown in Figure \ref{fig:h1}. We note this over-density is distinct from WLM's only known globular cluster, WLM-1, which is located on the western edge of WLM's outer disk \citep{Ables77, Sandage85, Hodge99}.

\begin{figure}
\includegraphics[width=\columnwidth]{"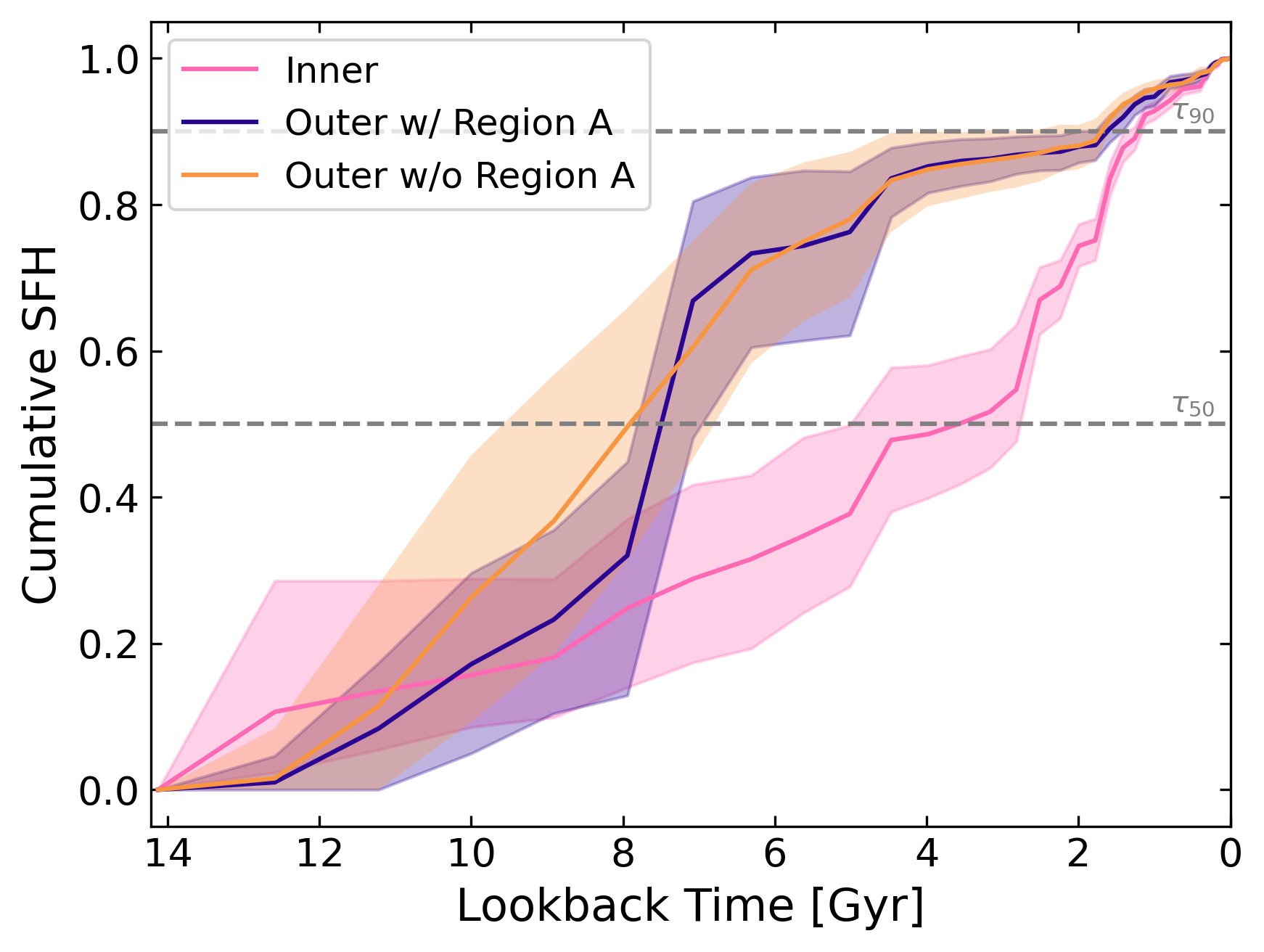"}
\includegraphics[width=\columnwidth]{"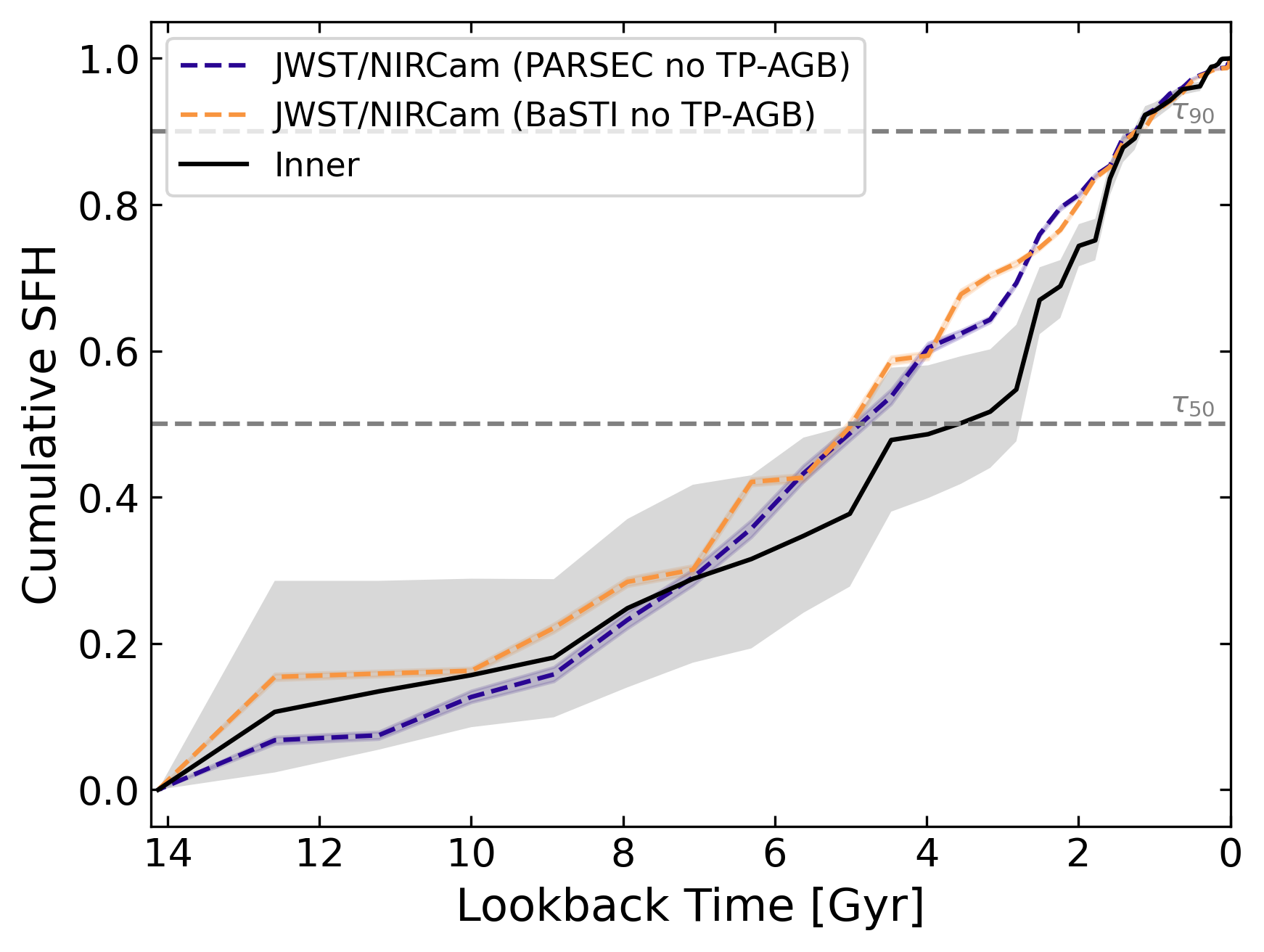"}
\includegraphics[width=\columnwidth]{"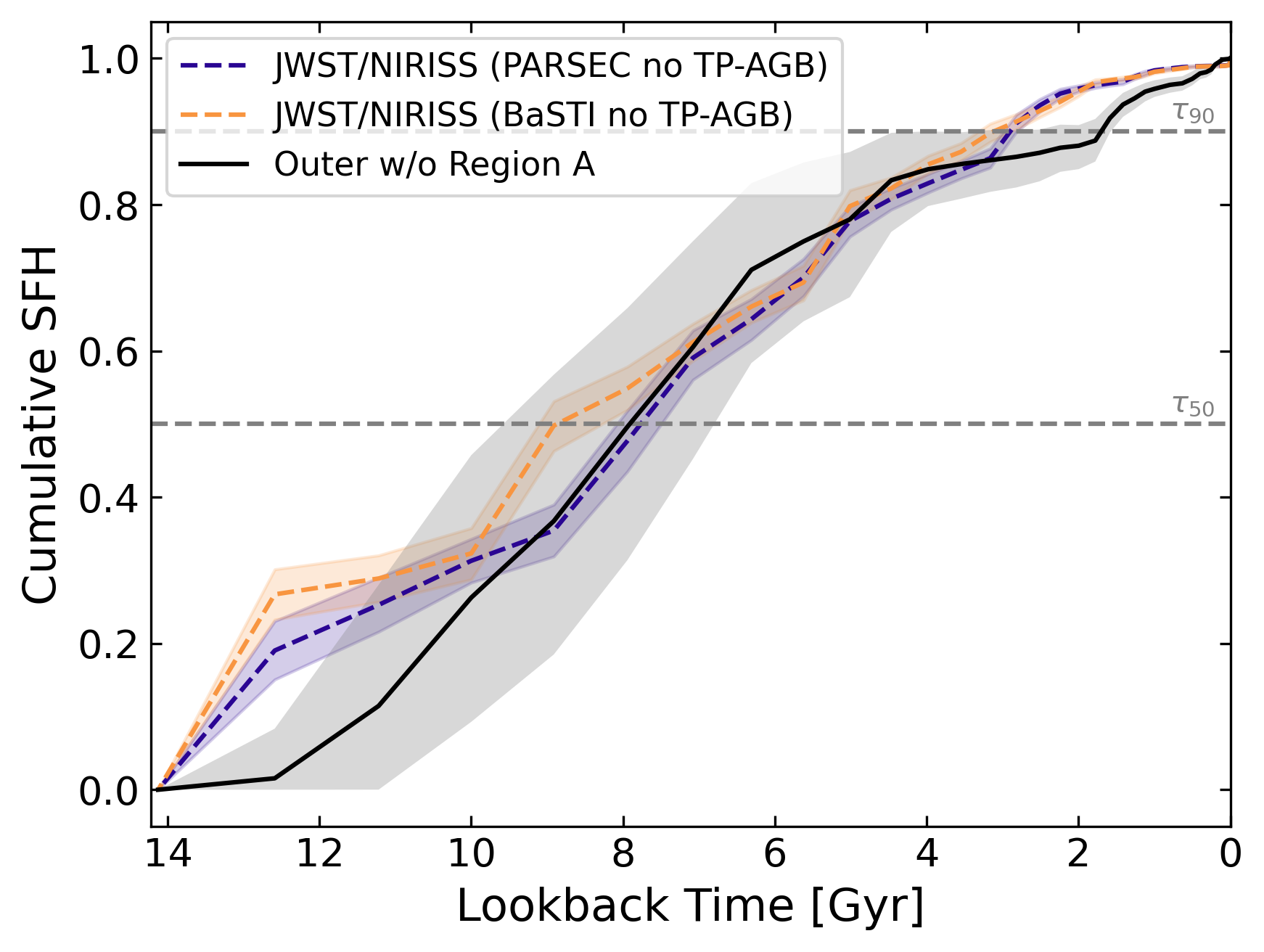"}
\caption{Top panel: cumulative star formation history for the inner ($r<r_h$), outer region ($r>r_h$) excluding Region A, \added{and outer region ($r>r_h$) including Region A}. Middle panel: cumulative star formation history for the inner region. The magenta and orange lines represent fits derived from the oMSTO using the PARSEC and BaSTI models, respectively, for the \citetalias{Cohen25} JWST/NIRCam field. Bottom panel: cumulative star formation history for the outer region ($r>r_h$), excluding Region A. The magenta and orange lines represent fits derived from the oMSTO using the PARSEC and BaSTI models, respectively, for the \citetalias{Cohen25} JWST/NIRISS field. } \label{fig:outer_burst} 
\end{figure}

\begin{figure}
\includegraphics[width=\columnwidth]{"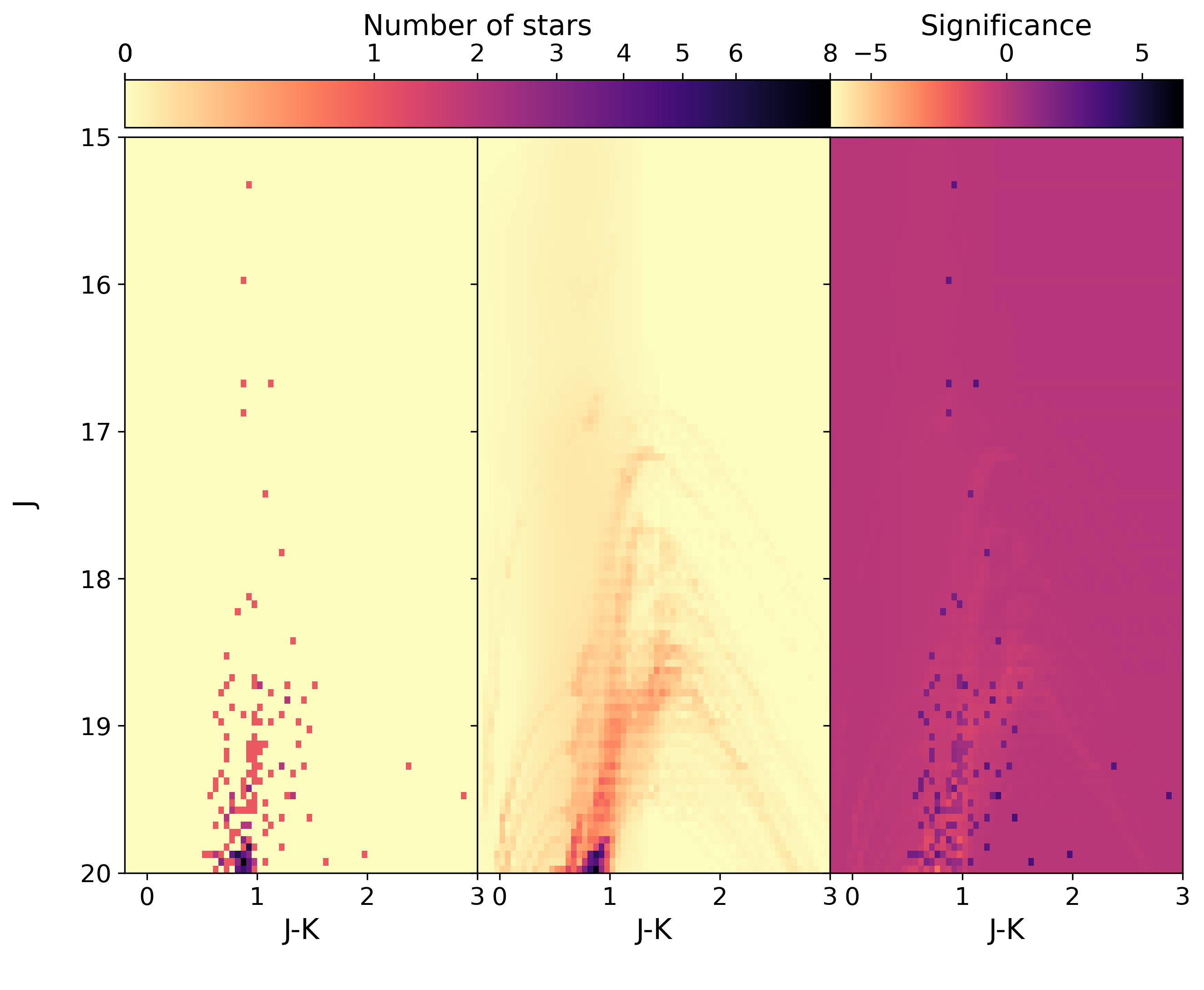"}
\caption{Hess diagram for the AGB star SFH fits for the observed data (left panel), best-fit linear combination of SSPs (middle panel), and residual CMD (right panel) for the outer region ($r>r_h)$ after the stars in Region A were removed. The residuals are expressed in Poisson standard deviations. } \label{fig:outer_residuals} 
\end{figure}

To characterize the stellar populations of this over-density, as well as its impact on our global SFH, we re-fit the SFH of the outer region ($r>r_h$) with Region A excluded. \
We show the resulting fit in Figure \ref{fig:outer_burst}.
\added{The residuals for this fit are shown in Figure \ref{fig:outer_residuals}.}
The burst at 8~Gyr seen in Figure \ref{fig:sfh} disappears. 
\added{To test whether this signal is real, we also removed a different set of stars that did not overlap with Region A, and  measured its SFH. We found the burst is still reproduced, demonstrating the burst does indeed result from Region A. This test is described in Appendix \ref{sec:appendixA}. }

\begin{figure*}
\gridline{\fig{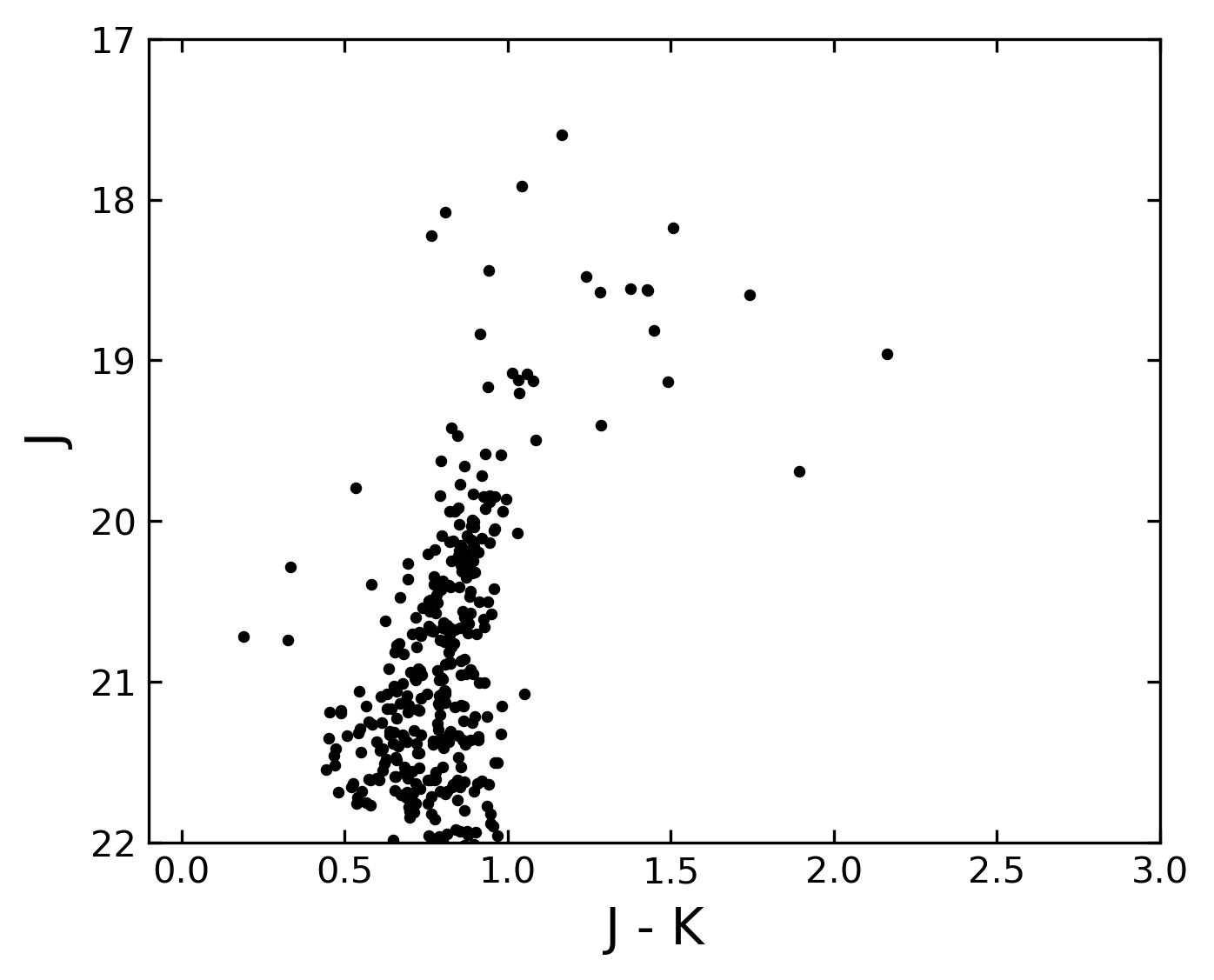}{0.5\textwidth}{}
\fig{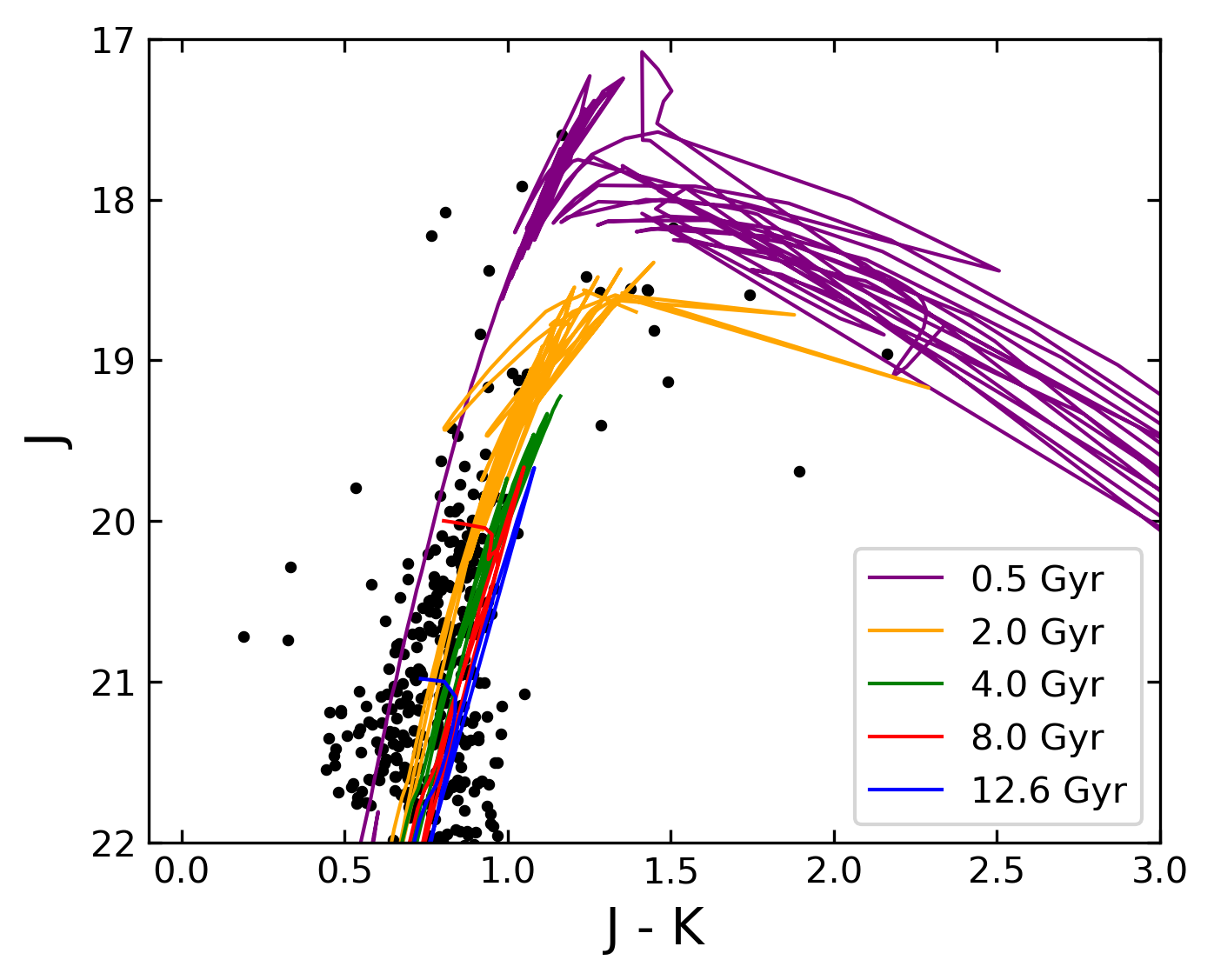}{0.5\textwidth}{}
          }
\gridline{\fig{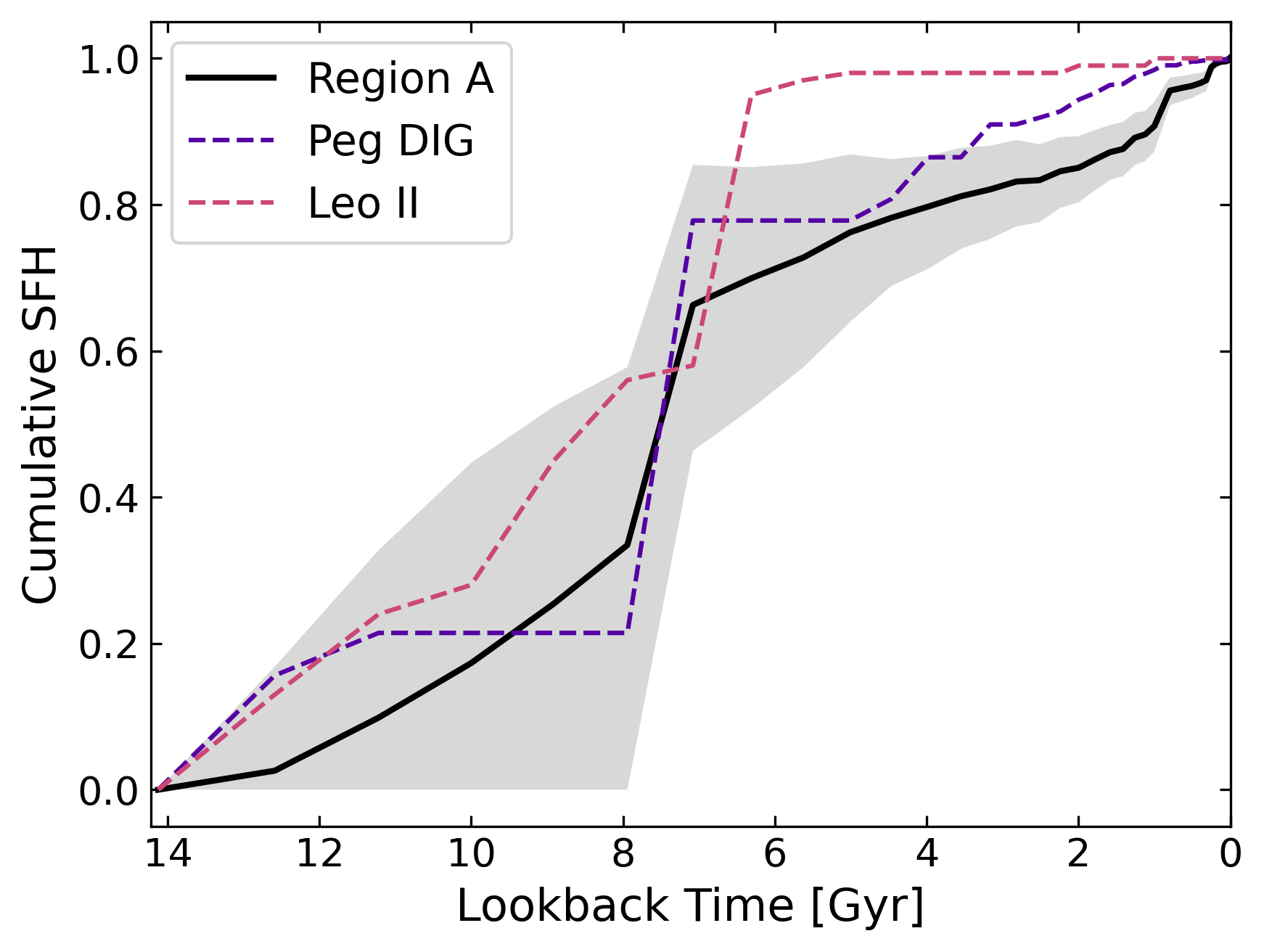}{0.5\textwidth}{}
\fig{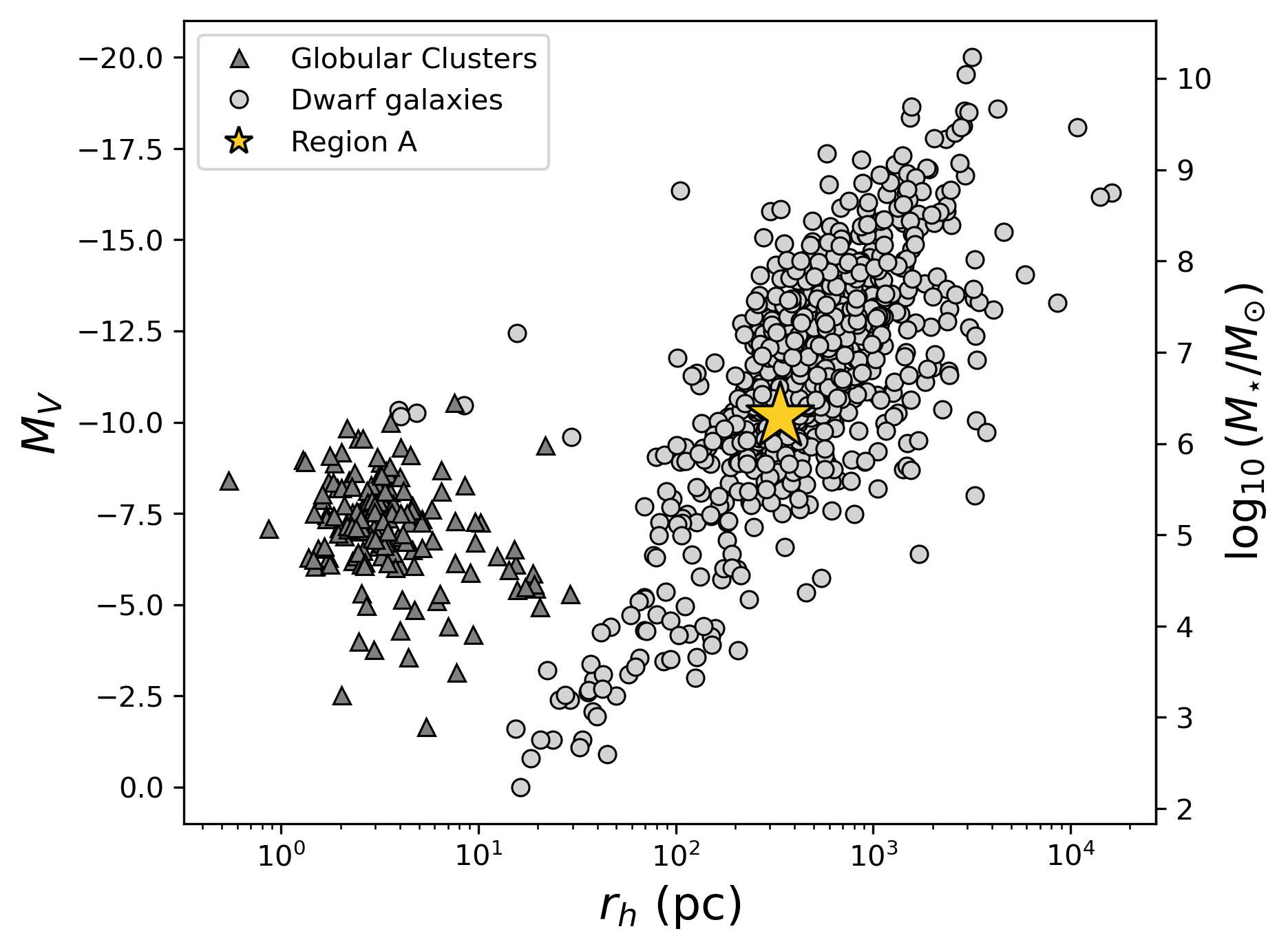}{0.5\textwidth}{}
          }
\caption{\added{Top panel: CMD of the stars in Region A. On the right, we show the same CMD overplotted with \texttt{COLIBRI} isochrones with a metallicity of $[M/H]=-1.0$~dex.} 
Bottom panel: (Left) Cumulative star formation history for Region A only. The spatial extent of Region A is shown in Figure \ref{fig:maps}.  Overplotted are SFHs for Pegasus and Leo II, two dwarf galaxies with similar masses and sizes to Region A \citep{Savino25, Weisz14b}.  (Right) Luminosity-size relationship for globular clusters and dwarf galaxies in the Local Volume (LV). Region A is denoted with a gold star. This figure has been modified from Figure 8 of \cite{Pace25}, who computed the masses assuming a mass-to-light ratio of 2. } \label{fig:dwarf} 
\end{figure*}

\added{We then fit the SFH of just Region A, whose CMD is shown in Figure \ref{fig:dwarf}}. We first removed the contribution of the outer region's intrinsic SF, by
adding the outer region's CMD as a `background CMD' to Region A's \texttt{MATCH} parameter file, scaled down by a factor of $\sim0.05$ (i.e., the angular area of the outer region divided by the angular area of Region A).
The resulting SFH fit for Region A only is shown in the middle panel of Figure \ref{fig:dwarf}.
The uncertainties are large, as expected, because the SFH is based on only 49 stars.  \added{In Appendix \ref{sec:appendixB}, we demonstrate through mock data that a $\sim8$~Gyr burst can be recovered to $\pm1.5$~Gyr from $\sim50$~AGB stars}

Qualitatively, there appears to be little-to-modest ancient star formation in this region with a rapid increase at $\sim8$~Gyr ago. This was followed by another period of roughly constant star formation. We calculated the total stellar mass of Region A to be $M_{*}=3.3\pm0.8\times10^6 ~M_{\odot}$ and that it formed 50\% of its stellar mass $\tau_{50}=7.5_{-0.9}^{+1.7}$~Gyr ago.  
We model the light profile of the over-density by fitting an exponential light profile (e.g., as described in \citealt{Martin08, Martin16}) to the AGB and RGB stars (down to the 80\% completeness level of the photometry).
\added{We computed the radial density across a grid of cells on and around Region A, and then fit the \cite{Martin16} radial density profile to this grid using \texttt{scipy.optimize.minimize} \citep{Scipy}}.
We find a half-light radius of $r_h=338\pm167$~pc, where the uncertainty was calculated using 1000 bootstrap realizations. 
The half-light profile is shown as a pink circle in Figure \ref{fig:image}. 
\added{We also found best-fit parameters for Region A of 
$\rm{R.A.}_0=0.481^{\circ}$, 
$\rm{Decl.}_0=-15.380^{\circ}$, and 
$\rm{P.A.}=-4.6^{\circ}$}.

We speculate on possible origins for this over-density of intermediate-age stars.  First, we consider the possibility that Region A is a star cluster that formed \textit{in situ} in WLM, possibly due to its location near the end of a reported bar in WLM \citep{Kolhe26}. This seems unlikely as Region A's radius of $338$~pc is significantly larger than the radii of even the most massive star clusters, which have half-light radii of $\sim100$~pc (e.g., \citealt{Krumholz19}). Given its large stellar mass and older age, it also seems unlikely to be a massive $\sim8$~Gyr cluster that is still in the process of dissolving, as most massive clusters dissolve on timescales of $\lesssim1$~Gyr (e.g., \citealt{Bica01, PortegiesZwart01,PortegiesZwart10, Kruijssen11}).

A second possibility is that this over-density is an accreted satellite galaxy.  
WLM has long been considered an archetypal `isolated' dwarf galaxy, and therefore an ideal testbed for understanding how low-mass galaxies evolve when unperturbed from external environmental effects, such as ram pressure or tidal effects (e.g., \citealt{Leaman09, Albers19, Sarajedini23, Archer24, McQuinn24a, Cohen25}).

\added{A largely intact dwarf galaxy within WLM has been predicted by simulations.  For example, this over-density could have been accreted $\sim8$~Gyr ago, triggering the burst of star formation in that area. Dynamical models of WLM from \cite{Leung21} and Leaman et al. (in preparation) show the location of Region A is   where WLM's tidal field transitions to a compressive mode, where any objects in that region would have an extremely long dissipation time ($\gtrsim$~Hubble time) and a core-stalling distance, which together would prevent the object from being destroyed  \citep{Petts15, Leung21}. Thus, Region A's projected distance explains how it may be a rare case of a non-tidally disrupted dwarf that is still intact \citep{Renaud09}.} Or, it could have been a dwarf galaxy accreted more recently, which would explain Region A's spatial association with H~{\sc i} disturbances \citep{Jackson04, Koch25}, which \cite{Khademi21} demonstrated may have been caused by a minor merger. 


Furthermore,  Region A's properties are similar to other nearby dwarf galaxies. We compare Region A's properties to other globular clusters and dwarf galaxies in the Local Volume, whose present-day masses (calculated from their $M_V$ values using a mass-to-light ratio of 2) and sizes have been compiled in \cite{Pace25}.  First, we convert Region A's CMD-based total stellar mass to a present-day stellar mass by accounting for stellar evolution mass loss effects, which \cite{Vincenzo16} estimates to be $\sim41\%$ for low-metallicity galaxies. Applying this correction to our total stellar mass yields a present-day mass of $M_{*,z=0}\approx2.0\times10^{6}~M_{\odot}$. 
In Figure \ref{fig:dwarf}, we plot Region A on the luminosity-size relation for globular clusters and dwarf galaxies in the Local Volume compiled in \cite{Pace25}. Furthermore, in Figure \ref{fig:dwarf}, we compare the cumulative SFH of Region A to the SFHs of Pegasus and Leo II \citep{Savino25, Weisz14b}, two dwarf galaxies with similar masses and sizes to Region A. Both these comparisons demonstrate that Region A's mass, size, and SFH are broadly similar to those of other dwarf galaxies in the Local Volume.


Further data, such as an oldest MSTO-based SFH which would decrease the SFH uncertainties and spectroscopic observations (e.g., kinematics, metallicities), would deliver a clearer picture of Region A's origin. 
More broadly, the discovery of this over-density, regardless of its origin, is another demonstration for discovery and characterization potential of  wide-area, NIR imaging of AGB stars in nearby galaxies. 

\section{Population Gradients in WLM}\label{sec:age_gradient}

\subsection{WLM's Age Gradient}
Comparing the SFHs of the inner ($r<r_h$) and outer ($r>r_h$, excluding Region A) regions in Figure \ref{fig:outer_burst} reveals a clear age gradient in WLM. We find the inner region formed 50\% of its stellar mass $\sim3.6$~Gyr ago, making it a fairly young region, whereas the outer region (without Region A) formed 50\% of its stellar mass $\sim7.9$~Gyr ago, making it older on average. 
 This age gradient is consistent with the `outside-in' radial stellar age gradient observed in most dwarf galaxies, where the stellar populations in the outskirts are on average older than in the central regions (e.g., \citealt{DohmPalmer98, Gallagher98, Lee99, Aparicio00, dejong08,delpino13, Hidalgo13, Mcquinn17, Savino19}). 
These age gradients are a result of stellar feedback, which drives stars initially formed in the central regions  to the outskirts 
 (e.g., \citealt{ElBadry16,Collin22}). 

We note we also find good agreement in the qualitative shapes of our inner and outer SFHs with the \citetalias{Cohen25} NIRCam-based SFH and NIRISS-based SFH shown in Figure \ref{fig:outer_burst}, where we assumed that the star formation of the NIRCam and NIRISS fields represent the star formation inside and outside of $1~r_h$, respectively. 
We also find good statistical agreement with \citetalias{Cohen25} in our measured values of $\tau_{90}$ and $\tau_{50}$ in the inner and outer regions, which are listed in Table \ref{tab:SFR_frac}.  

\subsection{WLM's Metallicity Gradient}
 While our study and others have found that WLM has a steep age gradient (e.g., \citealt{Albers19, Cohen25}), \cite{Leaman13} found WLM has a weak metallicity gradient of $-0.04~\rm{dex~kpc^{-1}}$ out to $\sim2 r_h$ on the major axis. 
Simulations have shown that galaxies with weaker metallicity gradients tend to be younger on average,
because recent, metal-rich star formation washes out pre-existing metallicity gradients caused by the radial migration of older, metal-poor stars \citep{ElBadry16, Mercado21}. 
This picture is consistent with our measured value of $\tau_{50}\approx5.2$~Gyr, which suggests WLM is a relatively young galaxy compared with other LG dwarf galaxies (e.g., \citealt{Weisz14b}), and is thus expected to have a weak metallicity gradient (e.g., \citealt{Taibi22, Fu24}).

\section{Future Prospects}\label{sec:future}

AGB star SFHs are well-suited for NIR observations with facilities such as Roman, Euclid, and JWST. Roman and Euclid can resolve AGB stars in the outer disks and halos of galaxies out to $\sim10$~Mpc. JWST can resolve AGB stars in the inner disks of galaxies out to $\sim10$~Mpc, and in the outer disks and halos of galaxies out to tens of Mpc.  Because they are significantly brighter than other features (e.g., AGB stars are $\sim100\times$ brighter than RC stars) similar SFH science can be achieved with AGB stars for less integration time. 

We illustrate the observational efficiency of AGB star SFHs by comparing hypothetical JWST observations of AGB stars and RC stars as a function of galaxy distance using the JWST exposure time calculator (ETC)\footnote{\url{https://jwst.etc.stsci.edu/}} \citep{Savino24}. These exposure times are listed in Table \ref{tab:exposure_time}.  We do not include crowding in this exercise, though note that it is far more likely to affect RC stars than AGB stars.
 
To simulate the observational time needed for the AGB star SFHs, we calculated the exposure time needed to resolve stars 1~mag below the TRGB at an SNR $= 5$ in the F115W filter in NIRCam's short wavelength (SW) channel (pixel scale: $0.031''~\rm{pixel^{-1}}$) and at a SNR of SNR $\approx7$ in the F277W filter in the long wavelength (LW) channel (pixel scale: $0.063''~\rm{pixel^{-1}}$). Measuring the color comes at no additional cost for JWST, which can observe simultaneously with the SW  and LW channels.  Even the LW channel on JWST is sufficiently sharp to resolve AGB stars at $\gtrsim40$~Mpc (e.g., \citealt{Li25}).  
Our tests suggest this is the optimal color combination for AGB star SFH measurements, \added{as it maximizes color separation on the CMD between different AGB star isochrones, while at the same time minimizing total exposure time}.  
These SNR requirements $\sim1$~mag below the TRGB are standard for datasets used to measure precise TRGB and JAGB distances with JWST (e.g., \citealt{Newman24, Freedman25}), which are well-matched for also measuring AGB star SFHs.

For RC SFHs, we adopt the filters F115W and F200W and assume that observations need to extend $\sim1$~mag below the RC.  The faintness of the RC increases the effects of crowding, making it necessary to use two SW filters, as opposed to one SW and one LW filter.  This particular filter combination has been prevalent in the JWST-based CMDs that target the RC  (e.g., \citealt{Habel24, Nally24,Bortolini25,Ck25,Correnti25}).  The need to reach $\sim1$~mag below the RC is motivated by the need to minimize systematics on the SFHs, which can become quite large if the entire RC is not included on the CMD \citep[e.g.,][]{Dolphin12, Weisz14b}.  Accordingly, we calculated the exposure time needed to resolve stars 1~mag below the median RC magnitude in the F115W filter at a color of $F115W-F200W<1$ (i.e., the blue edge of the RC) at an SNR $= 5$.

The results of this exercise are summarized in Table \ref{tab:exposure_time}.  They demonstrate that AGB star SFHs are $100\times$ more observationally efficient than RC-based SFHs, even without factoring in crowding effects. At $D\sim7$~Mpc, it only takes JWST 4 min to reach our target depth for AGB stars compared to 8 hours for the RC.  The difference is more pronounced at $D\sim13$~Mpc.  Beyond this distance, effectively the edge of the Local Volume, JWST requires prohibitively long integration times to resolve the entire RC.

This analysis is supported by JWST observations in the literature.
AGB stars have been resolved with NIRCam in the Virgo cluster ($d=16$~Mpc) in 18 min (e.g., \citealt{Anand25}), in disk galaxies at $D\sim21$~Mpc in 48~min (e.g., \citealt{Freedman25}). Brighter AGB stars ($M_J\lesssim-5$) have already been resolved in a galaxy at a distance of $d=40$~Mpc with JWST in 2~hours of exposure time \citep{Li25}.  Scaling from crowding and sensitivity estimates from these datasets, it is likely that JWST can resolve AGB stars in the halos of galaxies out to $D\gtrsim50$~Mpc in modest integration times (e.g., tens of hours).  Indeed, one ambitious JWST program is even targeting resolved AGB stars in the Coma Cluster ($D\sim100$~Mpc) by making use of JWST's extremely wide-pass filters (JWST-GO-6876, PI: W. Freedman), which, if successful, could extend resolved star-based SFHs to distances that have been previously unimaginable.

\begin{deluxetable*}{c|c|ccc}
\tablecaption{JWST/NIRCam exposure times required to resolve stellar populations at a $\rm{SNR}=5$}\label{tab:exposure_time}
\tablehead{
\colhead{Distance (Mpc)} & 
\colhead{Total AGB\tablenotemark{a}} & 
\colhead{RC magnitude\tablenotemark{b}} & 
\colhead{RC color\tablenotemark{c}}  &
\colhead{Total RC}}
\startdata
7 & 4 min & 4 hr & 4 hr & 8 hr\\
13 & 12 min & 79 hr & 70 hr & 149 hr \\
21 & 48 min & \nodata & \nodata & \nodata\\
40 & 5 hr & \nodata & \nodata & \nodata
\enddata
\tablenotetext{a}{1 mag below TRGB}
\tablenotetext{b}{1 mag below median RC magnitude}
\tablenotetext{c}{$F115W-F200W < 1$}
\end{deluxetable*}

\section{Conclusion}\label{sec:conclusion}

Using NIR observations of resolved AGB stars, we derived the lifetime SFH of the Local Group galaxy WLM.
We find that the SFH measured from AGB stars is in excellent agreement with the SFH measured from a deep CMD at the depth of the oMSTO. 
We also recover an age gradient in good agreement with the age gradient measured from deep JWST data, \added{where we measured the SFH of the outer ($r>r_h$) and inner regions ($r<r_h$) of WLM separately,} and found the outer region of is on average older than the inner region.

This test demonstrates that AGB stars can produce lifetime SFHs comparable \added{in accuracy} to those currently available in and around the Local Group, albeit with the potential to do so in galaxies at distances out to $\sim50$~Mpc.  Several ongoing efforts to improve AGB star models and age calibrations (e.g., \citealt{Boyer13, Ky25} and JWST-GO-6852; PI: S. Goldman) will continue to make AGB star SFHs even more powerful.
 
We also identified an over-density of AGB stars in the northwestern outer disk of WLM, associated with a burst of star formation $\sim8$~Gyr ago. We speculate it is likely an accreted satellite galaxy, given its size and inferred stellar mass ($r_h=338$~pc, $M_*=2.0\times10^6~M_{\odot}$).  Our characterization of this system is coarse as only $\sim50$ AGB stars are associated with it. A more precise SFH derived from a well-populated CMD at the depth of the oMSTO would significantly improve our understanding of this unusual feature.  Beyond revealing the intriguing hints of a low-mass past accretion event, this finding demonstrates the power of wide-field NIR imaging for the efficient discovery and characterization of substructure in galaxies.

\begin{acknowledgments}
\added{We sincerely thank the referee for their constructive comments which improved this paper. We thank Ryan Leaman for useful discussions.} We thank Roger Cohen for graciously supplying us with his star formation history fits. We also acknowledge Wendy Freedman, Barry Madore, and Andy Monson for their observing contributions. 

AJL is supported by NASA through the NASA Hubble Fellowship grant
\#HST-HF2-51580.001 awarded by the Space Telescope Science Institute, which is operated by the
Association of Universities for Research in Astronomy, Inc., for NASA, under contract
NAS5-26555.

This research has made use of NASA's Astrophysics Data System Bibliographic Services. This research has made use of the NASA/IPAC infrared Science Archive (IRSA), which is operated by the Jet Propulsion Laboratory, California Institute of Technology, under contract with the National Aeronautics and Space Administration \citep{irsa}. This paper includes data gathered with the 6.5 meter Magellan Telescopes located at Las Campanas Observatory, Chile. 
This paper uses data from the Local Group L-Band survey (LGLBS), which is an Extra Large program conducted on
Jansky Very Large Array, and is operated by the National Radio Astronomy
Observatory (NRAO) and includes observations from VLA projects 20A-346, 13A-213,
14A-235, 14B-088, 14B-212, 15A-175, 17B-162, and GBT projects 09A-017, 13A-420,
13A-430, 13B-169, 14A-367, 16A-413. NRAO is a facility of the National Science
Foundation operated under cooperative agreement by Associated Universities, Inc.
Execution of the LGLBS survey science was supported by NSF Award 2205628.

\end{acknowledgments}

\facilities{Magellan/FourStar}

\software{Astropy \citep{Astropy13, Astropy18, Astropy22}, \texttt{DAOPHOT} \citep{Stetson87, Stetson94, Stetson11}, \texttt{imf} (\url{https://github.com/keflavich/imf}), \texttt{MATCH} \citep{Dolphin02,Dolphin12,Dolphin13, Dolphin16}, Matplotlib \citep{Hunter07}, NumPy \citep{Harris20}, Pandas \citep{McKinney10}, Scikit-learn \citep{Pedregosa11}, Scipy \citep{Scipy}}

\appendix
\restartappendixnumbering

\section{Is Region A Responsible for the SF Burst in the Outer Region of WLM?}\label{sec:appendixA}

\begin{figure*}
\gridline{\fig{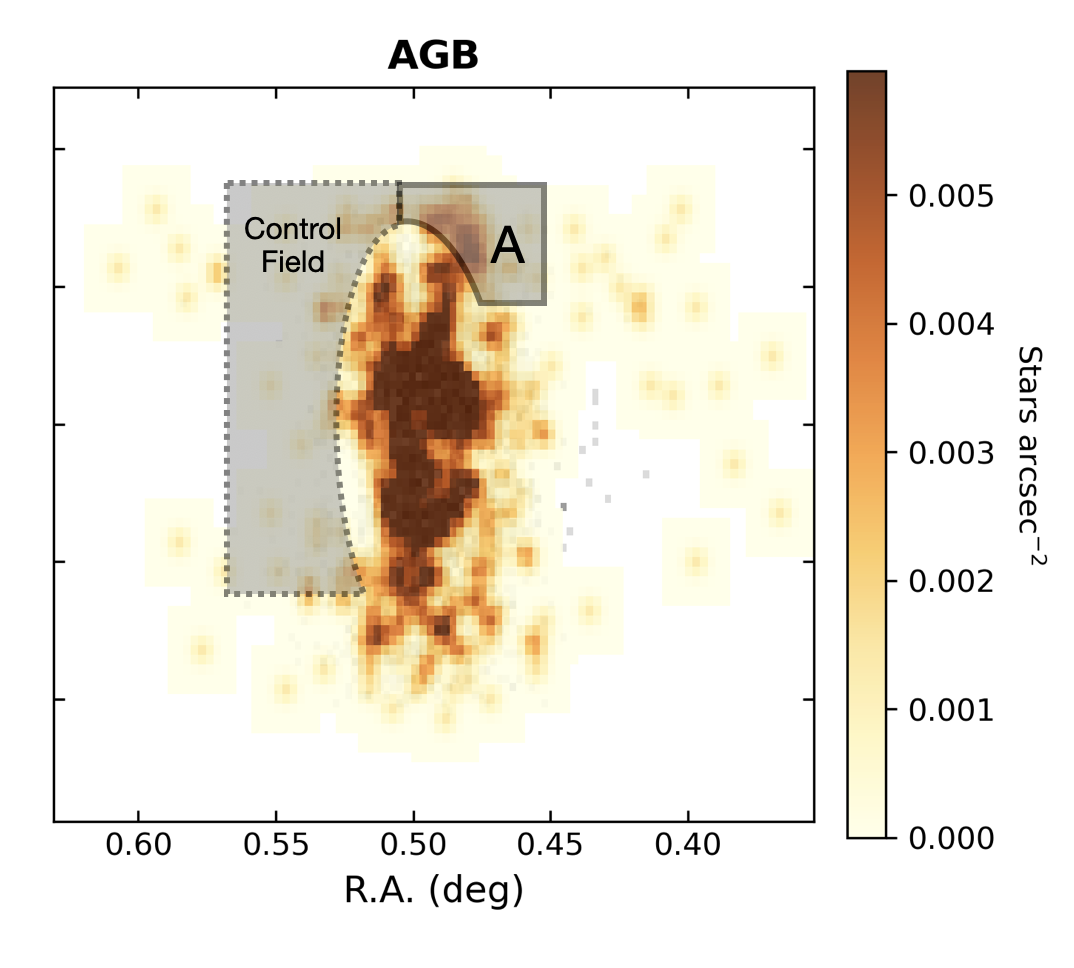}{0.5\textwidth}{}
\fig{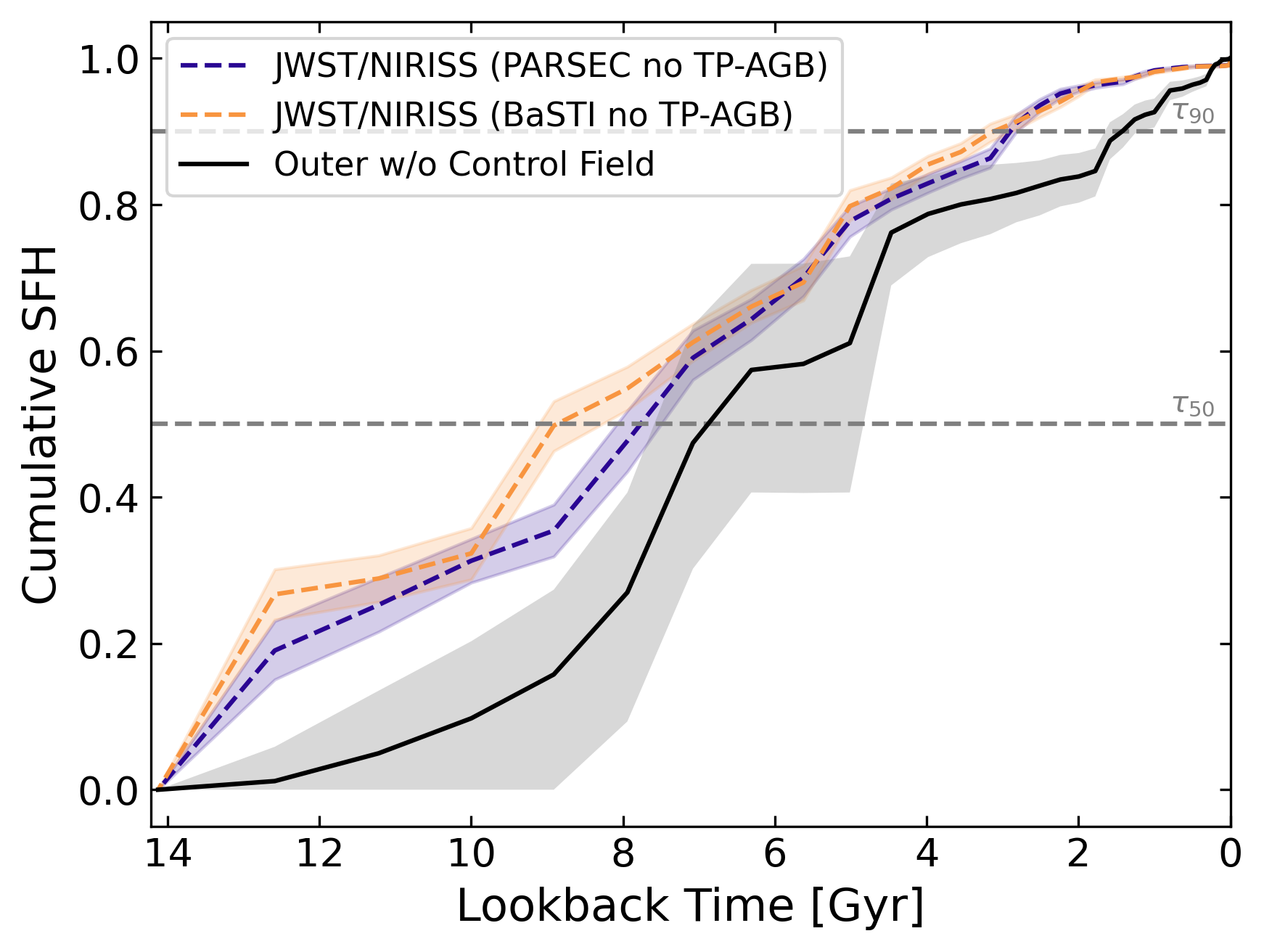}{0.5\textwidth}{}
          }
\caption{(Left panel) The field selected for this control test is denoted by the dotted grey-outlined region, while Region A is denoted by the solid grey-outlined region. Both are overlaid on an AGB stellar density map. (Right panel) The SFH for the outer region ($r>r_h$) with the control field removed, compared with the JWST/NIRISS SFHs.
}\label{fig:appendixA}
\end{figure*}

\added{In this section, we performed a control test to determine if Region A is responsible for the burst seen in the outer region's SFH shown in Figure \ref{fig:outer_inner}. We removed a set of 50~stars with $J<20$~mag that did not overlap with Region A  from the outer region. These stars were contained within the ``Control Field" shown in Figure \ref{fig:appendixA}. The angular area of the control field is larger than Region A, despite containing a similar number of AGB stars, because Region A contains an overdensity of AGB stars. We then re-measured the SFH of the outer region with the Control Field stars removed. The results of this exercise are shown in the right panel of Figure \ref{fig:appendixA}. 
The 8~Gyr burst is still recovered, unlike what is seen in Figure \ref{fig:outer_burst}, where the 8~Gyr burst completely disappears when the stars from Region A are removed. This test demonstrates the stars in Region A are indeed responsible for the 8~Gyr burst recovered in the outer region ($r>r_h$) SFH. 
}

\section{How well can you recover a SF Burst from $\sim50$~AGB stars?}\label{sec:appendixB}

\cite{Lee25} demonstrated $\gtrsim1000$~AGB stars are needed to recover a roughly constant star formation history through mock simulations. In this section, we test how well a SFH akin to Region A's can be recovered from $\sim50$~AGB stars.

\begin{figure*}
\gridline{\fig{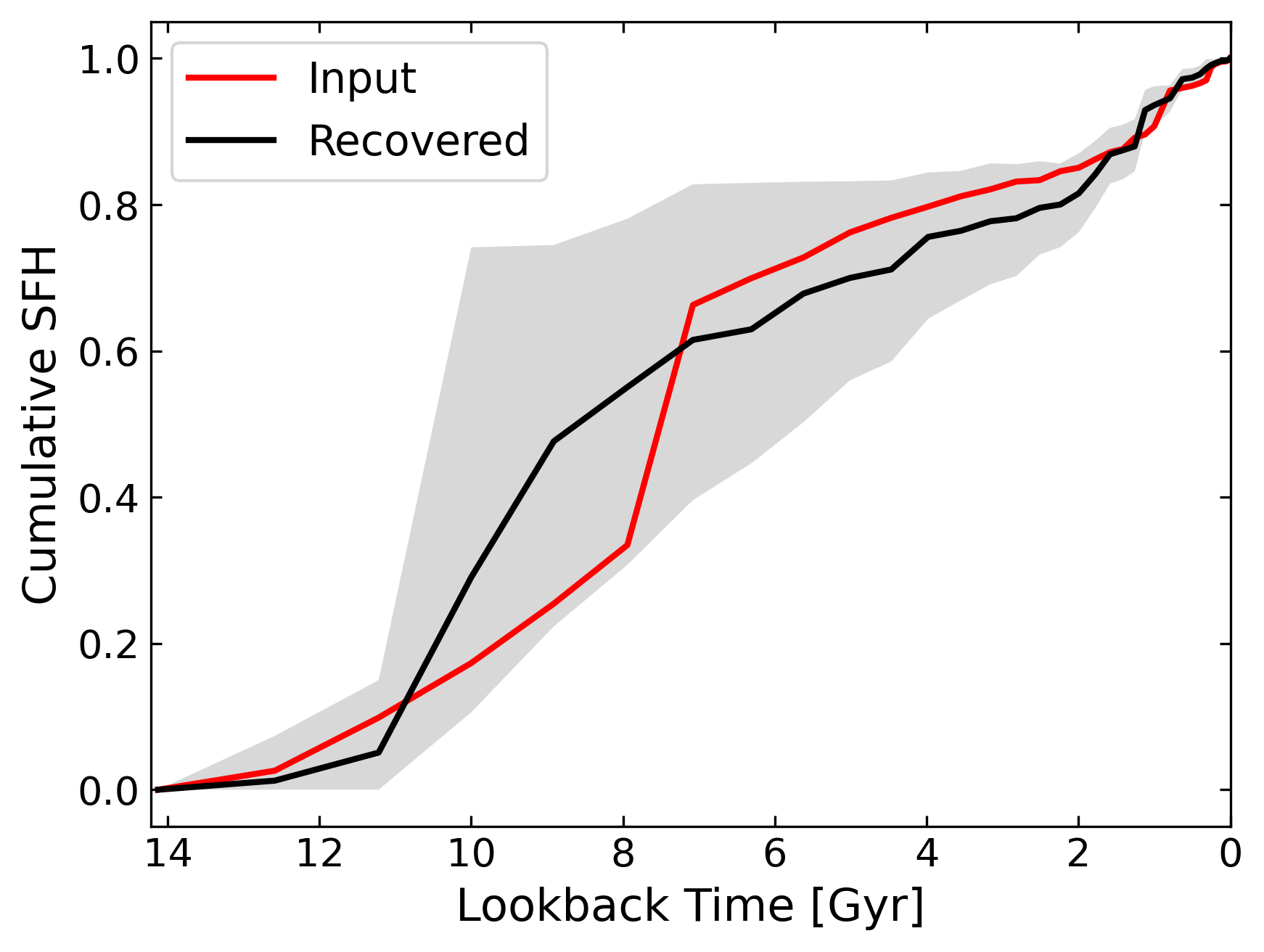}{0.5\textwidth}{}
\fig{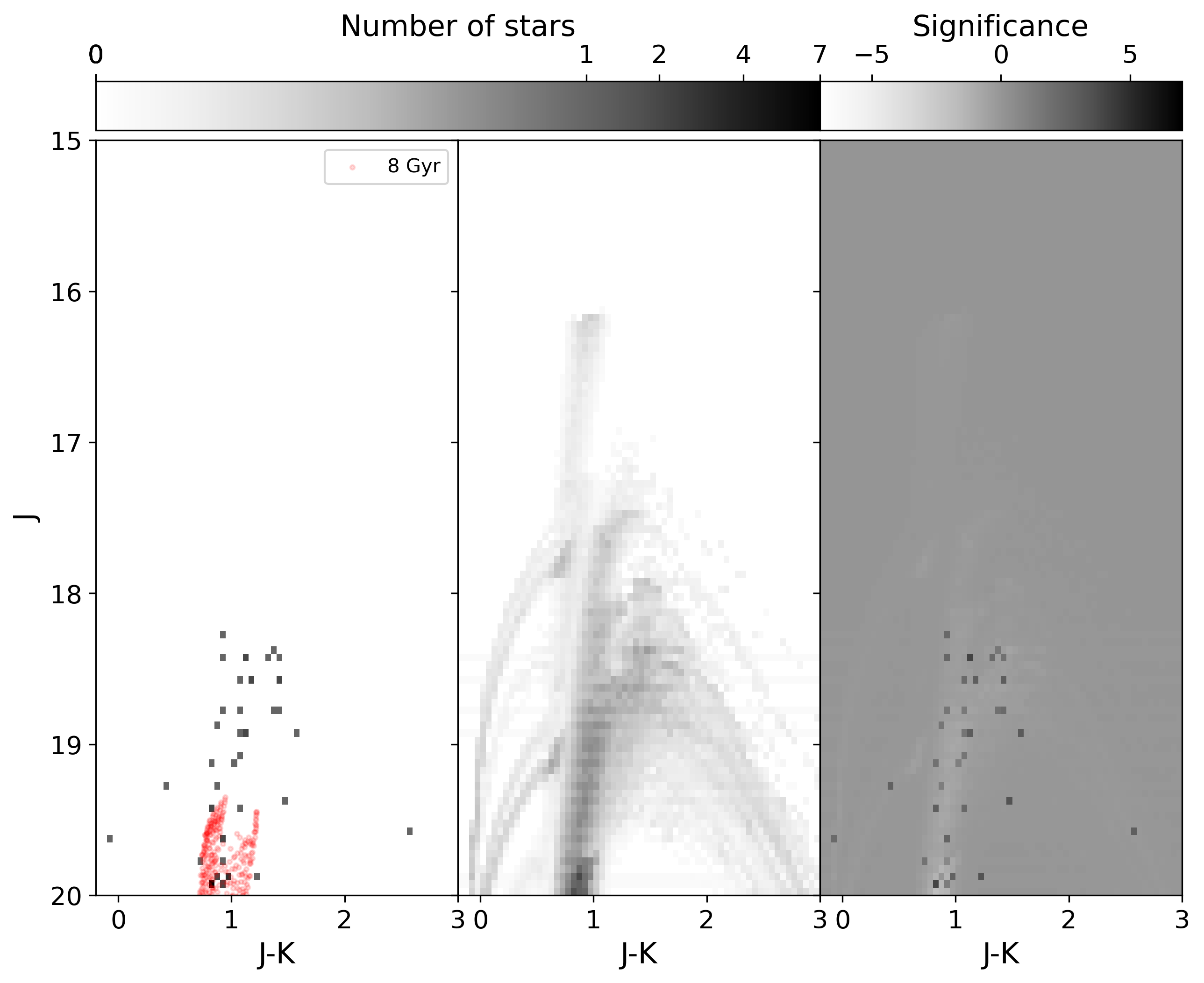}{0.5\textwidth}{}
          }\figurenum{B1}
\caption{(Left panel) Comparison between our injected star formation history and recovered star formation history. (Right panel) Hess diagrams from the result of this exercise for the mock observed data (left), best-fit linear combination of SSPs (middle) and residual CMDs (right). The mock observed CMD was generated using Region A's measured SFH. The residuals are expressed in Poisson standard deviations. In the left panel, the mock observed data are shown as black points.  The red points represent isochrones for stars with ages of 8~Gyr and metallicities ranging from $-2.0\le[M/H]\le-0.6$~dex in steps of 0.1~dex, demonstrating where the 8~Gyr age stars lie on this CMD.}\label{fig:appendixB}
\end{figure*}

We generated a synthetic CMD using the measured SFH of Region A as the input. This CMD contained $\sim50$~AGB stars with $J<20$~mag. We then ran this simulated CMD through \texttt{MATCH}. The results of this exercise are shown in Figure \ref{fig:appendixB}. We were able to recover the cumulative SFH to within its statistical uncertainties, demonstrating a SFH comprised of an $\sim8$~Gyr burst + roughly constant SFH can be recovered from $\sim50$~ AGB/RGB stars.

\bibliography{citations}{}
\bibliographystyle{aasjournalv7}

\end{document}